%% file: ms.tex
\pdfoutput=1
\documentclass[12pt]{article}
\usepackage{xcolor}
\usepackage{colortbl}
\definecolor{darkblue}{RGB}{0,0,139}
\usepackage[colorlinks,pdfpagelabels,pdfstartview=FitH,bookmarksopen=true,bookmarksnumbered=true,linkcolor=blue,plainpages=false,hypertexnames=false,citecolor=blue, urlcolor=blue]{hyperref}

\usepackage{amsfonts}
\usepackage{amsmath}
\usepackage{mathtools}
\usepackage{enumitem}
\usepackage{graphicx}
\usepackage{amssymb}
\usepackage{bm}
\usepackage{longtable}
\usepackage{authblk}
\usepackage{pdflscape}
\usepackage{rotating}
\usepackage[round]{natbib}
\usepackage{booktabs}
\usepackage{placeins}
\usepackage{siunitx}

\usepackage{setspace}
\usepackage{chngcntr}
\usepackage{verbatim}
\usepackage{multirow}
\usepackage{subcaption}
\usepackage{chngcntr}

\usepackage[title]{appendix}
\usepackage[gen]{eurosym}
\usepackage{comment}
\usepackage[bottom]{footmisc}
\usepackage{gensymb}
\usepackage{textcomp,mathcomp}
\usepackage{wrapfig}
\usepackage{pdflscape}
\usepackage{rotating}
\usepackage{epstopdf}
\usepackage{graphicx}
\usepackage{caption}
\usepackage{subcaption}
\usepackage{longtable,tabularx,ltxtable,ragged2e}
\usepackage{titlesec}
\newsavebox{\tablewidthbox}

\newcommand{\blind}{0}
\def\spacingset#1{\renewcommand{\baselinestretch}{#1}\small\normalsize
  \setlength{\abovedisplayskip}{6pt}
  \setlength{\belowdisplayskip}{6pt}
  \setlength{\abovedisplayshortskip}{4pt}
  \setlength{\belowdisplayshortskip}{4pt}}
\spacingset{1}

\titleformat{\section}
  {\normalfont\large\bfseries}{\thesection}{1em}{}
\titleformat{\subsection}
  {\normalfont\normalsize\bfseries}{\thesubsection}{1em}{}
\titleformat{\subsubsection}
  {\normalfont\normalsize\bfseries}{\thesubsubsection}{1em}{}

\begin{document}

\pagenumbering{arabic}
\if0\blind
\author[1]{Haroon Mumtaz}
\affil[1]{Queen Mary University of London}

\author[2]{Sofia Velasco}
\affil[2]{Banco de España}

\title{Nonlinear Drivers of Macroeconomic Tail Risk:\\ A Threshold Stochastic Volatility-in-Mean VAR with Regime-Dependent Leverage
\thanks{\footnotesize Corresponding author: Haroon Mumtaz (\href{mailto:h.mumtaz@qmul.ac.uk}{h.mumtaz@qmul.ac.uk}). The views expressed in this paper are those of the authors and do not represent those of Banco de España or the ESCB.}
}

\date{}
\maketitle
\fi
\if1\blind
  \begin{center}
    {\LARGE\bf Nonlinear Drivers of Macroeconomic Tail Risk: A Threshold Stochastic Volatility-in-Mean VAR with Regime-Dependent Leverage}
  \end{center}
\fi

\begin{abstract}
\setlength{\parskip}{2pt}
The tails of macroeconomic outcomes can respond differently from the centre of their distribution to aggregate shocks. We demonstrate this via a threshold stochastic-volatility-in-mean VAR with regime-dependent leverage. The model allows interactions between the endogenous variables and their volatility, contemporaneous level--volatility dependence, and regime-specific propagation. We estimate the model using a long-run U.S. data set, finding evidence for three inflation-based regimes. We identify business-cycle, financial, macroeconomic-uncertainty, and financial-uncertainty shocks and decompose their contributions to growth- and inflation-at-risk. Business-cycle shocks dominate the median response of GNP growth but account for a substantially smaller share of growth-at-risk. In contrast, uncertainty makes a material contribution to both growth- and inflation-at-risk.
\end{abstract}

\vspace{-1ex}

\clearpage
\spacingset{1}

\section{\label{intro}Introduction}

Recent contributions in macroeconomics and finance document pronounced asymmetries in the U.S.\ predictive distributions of economic growth and inflation. Financial conditions are particularly informative about downside risks to future activity \citep{Giglio2016,adrian2019,DelleMonacheDePolisPetrella2024}, while the determinants of inflation can affect parts of its predictive distribution differently, generating variation in upside inflation risk that is not captured by the conditional mean \citep{LopezSalidoLoria}. Moreover, \citet{loria-matthes-zhang-2025} show that monetary, financial,
uncertainty, and oil-price shocks can generate larger responses at the tenth
percentile of future activity than at the conditional median. The recurrence of this asymmetry across different shocks points towards a role for nonlinear propagation, rather than a mechanism specific to a single disturbance. This raises two related questions: do the shocks that drive conventional macroeconomic fluctuations also drive growth- and inflation-at-risk, and does their propagation through the predictive distribution depend on the state of the economy?

We address these questions by developing a threshold stochastic-volatility-in-mean VAR for the joint dynamics of GNP growth, inflation, the credit spread, and their endogenous stochastic volatilities. The model contains two interacting mean--volatility channels. First, past macroeconomic outcomes enter subsequent volatility dynamics, and innovations to macroeconomic outcomes and volatility can be contemporaneously correlated. Together, these features allow adverse innovations to be associated with higher subsequent volatility, a macroeconomic analogue of the leverage mechanism documented in financial markets \citep{Black1976,Christie1982,Schwert1989}. Second, stochastic volatility can feed back into expected macroeconomic
outcomes through a volatility-in-mean channel \citep{mumtazsurico2013,mumtaz2018,carriero2016,CrossHouKoopPoon}, consistent
with the broader real effects of uncertainty emphasised by \citet{Bloom2009}. The threshold structure allows these mean--volatility interactions, as well as the broader dynamics of the system, to change across inflation regimes. As a result, the same shock can reshape the joint predictive distribution of activity, inflation, and financial conditions differently depending on the prevailing inflation state.

The model combines the endogenous mean--volatility interactions of \citet{mumtaz2018} with the state-dependent transmission of uncertainty emphasised by \citet{alessandri2019financial} and \citet{CanovaForero24}. Most closely, \citet{CaldaraSVOL} use an SV-in-mean VAR to trace the
structural sources of macroeconomic and financial tail risk. Their model
generates state-dependent distributional responses through endogenous
mean--volatility interactions, but holds the parameters governing these
interactions, including the correlation between level and volatility innovations,
fixed over time. We allow
these coefficients and innovation correlations to change across regimes. The model can therefore distinguish
changes in the prevailing volatility state from changes in the parameters
governing mean--volatility feedback and shock transmission. Complementary data-rich evidence relates heterogeneous tail risk across variables to common mean--volatility interactions \citep{caldara2024risk}.\footnote{Related evidence separates common level and volatility factors in large macroeconomic panels and finds economically meaningful second-moment dynamics \citep{GorodnichenkoNg2017}. \citet{MumtazVelasco2026} model the joint evolution of level and volatility factors in a dynamic factor framework.}

Relatedly, \citet{ccm2024} show that BVARs with stochastic volatility can
capture more variable downside than upside growth risk even when their
one-step-ahead conditional predictive distributions are symmetric. Our focus
is on the structural drivers of these risks and how the parameters governing
their transmission vary across inflation regimes.

We apply the model to quarterly U.S.\ data from 1875 to 2024. The long sample permits the high-inflation state to be characterised from repeated episodes under distinct historical circumstances, rather than primarily from the Great Inflation. Predictive model selection based on recursive joint density forecasts of GNP growth and inflation supports a three-regime specification defined by inflation. 

The estimated inflation regimes characterise distinct propagation environments rather than volatility states. The Great Depression and the Global Financial Crisis feature exceptionally high innovation volatility but belong predominantly to the low-inflation regime. The reduced-form estimates indicate that the relationship between adverse output news and subsequent output volatility weakens as inflation rises.

The structural analysis delivers the paper's main result: the shocks that dominate conventional macroeconomic fluctuations need not dominate movements in the tails of the predictive distribution. Using max-share identification we show that the structural composition of tail risk differs from that of median responses. Business-cycle shocks dominate the median response of GNP growth but account for a substantially smaller share of growth-at-risk. Macroeconomic uncertainty makes a material contribution to both growth- and inflation-at-risk despite its limited role at the median. Its contribution to growth-at-risk rises with the magnitude of positive macroeconomic-uncertainty shocks, while the contribution of financial uncertainty to inflation-at-risk rises with the magnitude of financial-uncertainty shocks in high-inflation periods.

Section~\ref{model} presents the model and estimation. Section~\ref{dms} describes the data, model selection, risk measurement, and the estimated inflation regimes. Section~\ref{SVAR} identifies structural shocks, studies their nonlinear tail-risk responses, and decomposes the drivers of those responses. Section~\ref{conclu} concludes.

\section{\label{model}A Stochastic Volatility-in-Mean Threshold VAR with Leverage}

This section develops an empirical framework that combines the endogenous mean--volatility interaction of \citet{mumtaz2018} with the threshold-dependent transmission of uncertainty shocks in \citet{alessandri2019financial} and \citet{CanovaForero24}. The former allows level and volatility innovations to be contemporaneously correlated and volatility to feed back into macroeconomic outcomes; the latter shows that the transmission of uncertainty can vary across regimes. We allow these two dimensions to interact: both the coefficients governing macroeconomic and volatility dynamics and the correlations between level and volatility shocks can change across regimes. Macroeconomic risk can therefore vary not only with the level of volatility, but also with how level and volatility shocks comove and propagate. The regimes are determined by an observed lagged variable, while the number of regimes and the variable defining them are selected by out-of-sample predictive performance in Section~\ref{dms}.

\subsection{Model specification}
The regime-switching VAR model with stochastic volatility describes the log-volatility processes and the associated reduced-form dynamics of the VAR system:
\begin{align}
\tilde h_{t+1}&=\sum_{m=1}^{M}\Big(\alpha_m+\theta_m \tilde h_t+\sum_{j=1}^{Q}d_{j,m}\,Y_{t-j}\Big)\mathcal{I}(S_t=m)+\sum_{m=1}^{M}\tilde S_m^{1/2}\,\mathcal{I}(S_t=m)\,\eta_t, \label{eq:volh}\\
Y_t&=\sum_{m=1}^{M}\Big(c_m+\sum_{j=1}^{P}\beta_{j,m}\,Y_{t-j}+\sum_{k=1}^{K}b_{k,m}\,\tilde h_{t-k}\Big)\mathcal{I}(S_t=m)+H_t^{1/2}e_t. \label{eq:obs}
\end{align}
Equation~(\ref{eq:volh}) describes the evolution of the stochastic volatilities, collected in the $N\times1$ vector of unobserved log-volatility processes $\tilde h_t$. Their dynamics depend on lagged values of $\tilde h_t$ through $\theta_m$, which is allowed to be a full matrix and therefore permits the volatility processes to interact dynamically with one another. Volatility is also endogenous: through $d_{j,m}$, past values of the observable variables affect the evolution of $\tilde h_{t+1}$. The innovations to the volatility equation are scaled by $\tilde S_m^{1/2}$, where $\tilde S_m=\mathrm{diag}(\tilde s_m)$ and $\tilde s_m=[s_{1,m},s_{2,m},\dots,s_{N,m}]'$. Thus, both the coefficients governing volatility dynamics and the variances of the volatility shocks are regime specific, with $S_t\in\{1,\dots,M\}$ denoting the regime prevailing at time $t$ and $\mathcal{I}(S_t=m)$ the corresponding indicator.

\noindent Equation~(\ref{eq:obs}) describes the evolution of $Y_t$, the $N\times1$ vector of observable variables. Its conditional mean depends on lags of $Y_t$ through the coefficients $\beta_{j,m}$ and on lags of $\tilde h_t$ through $b_{k,m}$. The latter introduces the stochastic-volatility-in-mean channel: changes in volatility can therefore affect expected macroeconomic outcomes rather than only their dispersion. The reduced-form innovation is $H_t^{1/2}e_t$, where $e_t$ is an $N\times1$ vector of standardised residuals and $H_t=\mathrm{diag}(\exp(\tilde h_t))$ collects their time-varying variances. As in the volatility-transition equation, all reduced-form coefficients are allowed to vary across regimes.

The disturbances are jointly distributed as
\begin{equation}
\varepsilon_t=
\begin{pmatrix}
\underbrace{\eta_t}_{N\times1}\\[10pt]
\underbrace{e_t}_{N\times1}
\end{pmatrix}
\sim N(0,\Sigma_m),
\qquad
\underbrace{\Sigma_m}_{2N\times2N}=
\begin{pmatrix}
\Sigma_{\eta,(m)} & \Sigma_{\eta e,(m)}\\[2pt]
\Sigma_{\eta e,(m)}' & \Sigma_{e,(m)}
\end{pmatrix},
\label{eq:sigmar}
\end{equation}
where each block $\Sigma_{\eta,(m)}$, $\Sigma_{e,(m)}$, and $\Sigma_{\eta e,(m)}$ is $N\times N$. The diagonal elements of $\Sigma_m$ are restricted to one. The off-diagonal block $\Sigma_{\eta e,(m)}$ captures the contemporaneous correlation between shocks to the stochastic volatilities and shocks to the endogenous variables: its $(i,j)$ element is $\operatorname{corr}(\eta_{i,t},e_{j,t})$. Because $\Sigma_{\eta e,(m)}$ is regime specific, both the sign and the strength of this leverage relationship can vary across regimes. $\Sigma_{\eta e,(m)}$ describes the contemporaneous correlation between the level and volatility innovations, while the volatility innovation affects the subsequent log-volatility state. Restricting the diagonal of $\Sigma_m$ to unity makes it a correlation matrix, so that all scale is carried by $\tilde S_m^{1/2}$ and $H_t^{1/2}$. The resulting time-varying covariance matrix of the reduced-form disturbances in equations~\eqref{eq:volh}--\eqref{eq:obs} is

\begin{equation}
\underbrace{\Omega_t}_{2N\times2N}=
\sum_{m=1}^{M}\mathcal{I}(S_t=m)
\begin{pmatrix}
\tilde S_m^{1/2} & 0\\
0 & H_t^{1/2}
\end{pmatrix}
\Sigma_m
\begin{pmatrix}
\tilde S_m^{1/2} & 0\\
0 & H_t^{1/2}
\end{pmatrix}',
\label{eq:leverage}
\end{equation}

which combines regime-specific shock correlations and volatility-shock variances with the continuously evolving stochastic volatilities in $H_t$.

The regime process is defined by
\begin{align}
S_t&=1 &&\text{if } Y^{*}_{t-\mathrm{delay}}\le \mathrm{tar}_1,\notag\\
S_t&=m &&\text{if } \mathrm{tar}_{m-1}< Y^{*}_{t-\mathrm{delay}}\le \mathrm{tar}_m, \quad m=2,\dots,M-1,\notag\\
S_t&=M &&\text{if } Y^{*}_{t-\mathrm{delay}}> \mathrm{tar}_{M-1}, \label{eq:regime}
\end{align}
where $Y^{*}_t$ is a threshold series, a possibly transformed lag of one of the endogenous variables; $\mathrm{delay}$ is the lag at which it enters; and $\mathrm{tar}_1\le\dots\le\mathrm{tar}_{M-1}$ are the $M-1$ ordered thresholds. Because the regime is determined by a lagged endogenous variable, regime switches are endogenous rather than governed by a latent transition process.\footnote{The threshold variable and the number of regimes $M$ are selected across candidate specifications in the out-of-sample exercise of Section~\ref{dms}. Conditional on a specification, the threshold values and delay are estimated within the model.}

Following \citet{CaldaraSVOL}, the model contains an endogenous-volatility, or leverage, mechanism and a distinct volatility-in-mean mechanism. The leverage mechanism has a dynamic component, through which past macroeconomic outcomes affect subsequent volatility via $d_{j,m}$, and an innovation component, through which level innovations are correlated with innovations to subsequent volatility via $\Sigma_{\eta e,(m)}$ \citep{Black1976,Christie1982,Schwert1989}. The latter is the regime-specific leverage correlation reported below. Separately, volatility feeds back into expected macroeconomic outcomes through $b_{k,m}$ \citep{mumtaz2018}, consistent with the broader real effects of uncertainty emphasised by \citet{Bloom2009}. The threshold structure allows all three components, together with the remaining macroeconomic and volatility dynamics, to vary across regimes.

\subsection{\label{mc}Estimation}

We estimate the model using a Gibbs sampling algorithm. This section describes the prior distributions and summarises the main blocks of the sampler; full derivations and implementation details are provided in the Appendix.\footnote{We assess finite-sample recovery in calibrated two- and three-regime simulations with regime-specific dynamics and leverage correlations. In both cases, the sampler accurately recovers the thresholds and delay, tracks the latent log-volatilities closely, and produces coefficient credible intervals that cover the true values in most cases. The full design and results are reported in the Appendix.}

\subsubsection{Priors and starting values}

\paragraph{VAR and volatility-transition coefficients} Let $\Gamma_m$ collect the coefficients of the observation equation~\eqref{eq:obs} in regime $m$, and $\tilde\Gamma_m$ those of the volatility-transition equation~\eqref{eq:volh}. For both, we follow \citet{Banbura-Giannone-Reichlin-10Paper} and implement a Minnesota-type prior through dummy observations, with a common overall tightness of $0.2$ and prior means for each variable's own first-lag coefficient obtained from individual AR(1) regressions on a pre-sample training period. The coefficients on lagged stochastic volatilities in
equation~\eqref{eq:obs} use a separate prior scale,
$\tilde c=1$, in the dummy-observation notation of the
Appendix, while the prior on every intercept is set effectively flat. Because these dummy observations discipline only the coefficients and not the regime- and time-varying error covariance, the remaining parameters are assigned priors separately below.

\paragraph{Initial log-volatility} Following \citet{cogley-sargent-05}, we use a pre-sample training period to set a Gaussian prior for the initial state $\tilde h_0$, centred on the logarithm of the diagonal elements of the OLS residual covariance matrix estimated on that pre-sample, with prior variance $0.1I$.

\paragraph{Volatility-shock variances and the correlation matrix $\Sigma_m$} The diagonal elements of $\tilde S_m$ are assigned independent inverse-Gamma priors. For $\Sigma_m$, the unit-diagonal restriction already imposed in equation~\eqref{eq:sigmar} leaves only its off-diagonal correlations free; for these we assume a flat prior over the region in which $\Sigma_m$ is positive definite, the natural counterpart of the ``separation strategy'' of \citet{44b58bef-fe61-3589-95e9-7e6440f4511a}, which factors a covariance matrix into standard deviations and a correlation matrix. Unlike a prior placed on the elements of a Cholesky factorisation of $\Sigma_m$, this flat prior is invariant to how the shocks $\varepsilon_t$ are ordered.

\paragraph{Thresholds and delay} Each threshold is assigned a Gaussian prior centred on an economically meaningful percentile of the candidate threshold variable: for the benchmark inflation-threshold specification, the lower threshold is centred on the $50$th percentile and the upper on the $80$th, with a common prior variance of $0.1$, tight relative to the range of every candidate threshold variable. Draws that violate the ordering $\mathrm{tar}_1\le\dots\le\mathrm{tar}_{M-1}$ or leave any regime with too few observations receive zero prior mass. The delay has a discrete uniform prior over its admissible values; as noted above, the threshold variable itself is chosen across candidate specifications by the out-of-sample exercise of Section~\ref{dms}, rather than sampled within a given specification.

\subsubsection{Posterior simulation}

Let $\Psi$ denote the remaining parameters and latent states. The Gibbs sampler cycles through the following blocks:

\begin{enumerate}[itemsep=3pt,parsep=0pt,topsep=4pt]

\item \textbf{Thresholds and delay.}
Conditional on the remaining parameters, the threshold posterior is proportional to the likelihood times the prior, subject to the ordering and minimum-regime-size restrictions. Because the likelihood is a step function of the thresholds, we sample them using a shrinkage slice sampler. The delay is drawn from its discrete conditional posterior, with probabilities proportional to the likelihood at each admissible value.

\item \textbf{Regime-specific coefficients.}
Given the thresholds and delay, the regime allocation $S_t$ is known. Conditional on the stochastic volatilities and the remaining covariance parameters, the coefficients in equations~\eqref{eq:volh} and~\eqref{eq:obs} are sampled regime by regime from conditionally Gaussian regression posteriors. Draws implying explosive dynamics are rejected.

\item \textbf{Volatility-shock variances.}
Conditional on the level and volatility innovations, the diagonal elements of $\tilde S_m$ have nonstandard conditional posteriors because the level and volatility disturbances are correlated. We sample these parameters using an independence Metropolis--Hastings step based on an inverse-Gamma approximation to the conditional posterior.

\item \textbf{Regime-specific correlation matrices.}
Given standardised level and volatility innovations, each $\Sigma_m$ is sampled over the space of positive-definite correlation matrices. We update its free correlations one at a time, in random order, using slice sampling; the positive-definiteness restriction implies an admissible interval for each conditional update \citep{44b58bef-fe61-3589-95e9-7e6440f4511a}. This avoids tuning a random-walk proposal and enforces the
positive-definiteness restriction within each scalar update.

\item \textbf{Stochastic volatilities.}
Conditional on the remaining parameters, equations~\eqref{eq:volh}--\eqref{eq:obs} define a nonlinear state-space system because the latent volatilities enter both the conditional mean and variance of $Y_t$. We draw the full path $\{\tilde h_t\}_{t=1}^{T}$ using particle Gibbs with ancestor sampling \citep{JMLR:v15:lindsten14a}, with the state-space matrices switching according to the current regime allocation.

\end{enumerate}

\section{\label{dms}Regime-Dependent Leverage and Macroeconomic Tail Risk}
Our empirical analysis uses a parsimonious three-variable model and nearly 150 years of U.S.\ quarterly data. We compare candidate threshold variables and numbers of regimes by the fit of their recursive joint predictive densities to realised GNP growth and inflation. On this criterion, the data favour the specification with three regimes defined by annual inflation,\footnote{Throughout, annual inflation in quarter $t$ denotes the sum of the quarterly inflation rates over quarters $t-3$ to $t$, rather than a calendar-year average; annual real GNP growth is defined analogously.} which we use as the benchmark in the reduced-form and structural analyses below. Throughout the empirical analysis, we define growth-at-risk as the 5th percentile of the predictive distribution of GNP growth and inflation-at-risk as the 95th percentile of the predictive distribution of inflation.

\subsection{Data and variables}

The raw quarterly data span 1875Q2--2024Q1. After transformations, lags, and the pre-sample used to calibrate the priors, the estimation sample begins in 1881Q3. The three variables are real GNP growth, inflation, and the corporate credit spread. Real activity is measured by the quarterly growth rate of real GNP, inflation by the quarterly growth rate of the GNP deflator, and the credit spread by the difference between a corporate bond yield and the 10-year U.S.\ government bond yield. This three-variable macro-financial set-up closely follows \citet{mumtaz2018} and \citet{CaldaraSVOL}, facilitating comparison with their single-regime frameworks.

The historical series combine pre-1947 data from the NBER tables accompanying \emph{The American Business Cycle} with postwar data from FRED and the Global Financial Database. Real GNP and the GNP deflator are each spliced at 1947, while the corporate bond yield combines the NBER series with Moody's Baa yield. The 10-year government bond yield is obtained from the Global Financial Database over the full sample. The Appendix lists the underlying series and their identifiers and provides details on the transformations and splicing.

The length of the sample is particularly important for the threshold exercise because, unlike postwar samples, it contains repeated high-inflation episodes. These include World War I, the 1946--48 inflation surge, the Great Inflation, and the 2022--23 post-pandemic episode. This repeated variation allows us to assess whether changes in mean--volatility interactions recur systematically across high-inflation environments rather than reflecting a particular historical episode. This distinguishes our application from \citet{CaldaraSVOL}, whose estimation sample, 1954Q2--2019Q4, contains the Great Inflation but neither the earlier inflation episodes nor the post-pandemic surge.

\subsection{Model selection}

We compare seven specifications to determine which variable defines the regimes and how many regimes are required. The no-threshold model serves as the reference specification. The remaining specifications combine two or three regimes with thresholds in annual real GNP growth, annual inflation, or the credit spread.

Following \citet{GEWEKE2010216}, we compare specifications using joint one-quarter-ahead log predictive scores. At each forecast origin, we re-estimate each model on the expanding information set $\mathcal{D}_t$ and score the posterior predictive density of $Z_{t+1}$, the realised pair of GNP growth and inflation. The criterion rewards models that assign greater probability to the realised joint outcome while accounting for predictive dependence and uncertainty about parameters and latent states. For specification $\mathcal{M}_j$, the cumulative log predictive score is
\begin{equation}
\mathrm{LS}_j=\sum_{t=t_0}^{t_1}\log p\!\left(Z_{t+1}\mid\mathcal{D}_t,\mathcal{M}_j\right),
\label{eq:modelselection_ls}
\end{equation}
where $t_0=1975\mathrm{Q}1$ and $t_1=2022\mathrm{Q}1$, giving 189 common forecast origins. The joint predictive density is approximated by a bivariate kernel density estimate of the posterior predictive simulations. Summing its logarithm over the forecast origins yields a measure of each specification's sequential predictive fit to the sequence of realised GNP-growth--inflation pairs. Differences in cumulative scores measure relative sequential predictive fit for the target pair over the evaluation sample.

Table~\ref{tab:modelselection} reports the resulting scores. Among the seven specifications considered, the specification with three regimes defined by annual inflation attains the highest cumulative score. Under this one-quarter-ahead joint criterion, we select it as the benchmark specification for the reduced-form and structural analyses below. This selection also supports the argument in \cite{CanovaForero24} that the distribution of U.S.\ inflation is non-Gaussian and can be approximated via different regimes.

\input{tables/tab_modelselection}

\subsection{\label{post}Inflation regimes}

\begin{figure}[!htbp]
\centering
\includegraphics[width=\textwidth]{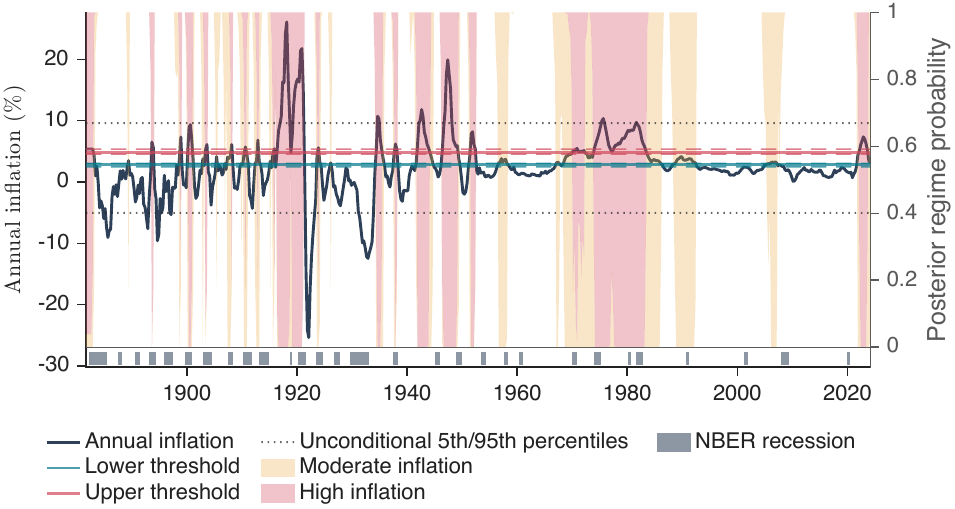}
\caption{\label{fig:regimeprob}Inflation regimes and posterior threshold estimates, 1881Q3--2024Q1.}
\caption*{\footnotesize \textbf{Notes:} The dark line is the threshold variable, annual inflation, at the posterior modal delay (left axis). Solid horizontal lines report posterior median estimates of the lower and upper thresholds; dashed lines show the corresponding 90\% credible intervals. The dotted horizontal lines are the unconditional 5th and 95th percentiles of annual inflation over the estimation sample, $-5.05$ and $9.61$ percent. Shaded areas report the posterior probabilities of the moderate- and high-inflation regimes (right axis); quarters in the low-inflation regime are left unshaded. Grey bars in the lane below the panel mark NBER recessions, from the peak to the trough quarter.}
\end{figure}

We re-estimate the selected three-regime specification, in which annual inflation defines the regimes, over the full estimation sample. Figure~\ref{fig:regimeprob} reports the threshold variable, posterior estimates of the two thresholds, and the associated regime probabilities. The posterior median thresholds are $2.86$ and $4.84$ percent, with 90\% credible intervals of $[2.47,3.05]$ and $[4.54,5.37]$, respectively. On average, the posterior regime probabilities assign $61\%$, $17\%$, and $21\%$ of the sample to the low-, moderate-, and high-inflation regimes. \footnote{ Over a postwar sample, \cite{CanovaForero24} report a threshold of 5.3 percent annual inflation in their model with the assumption of two regimes. Their high-inflation regime over this period closely matches our estimates of regime $3$. }

The estimated regimes have a clear historical interpretation. The high-inflation regime encompasses World War I, the wartime and postwar inflations of the 1940s and early 1950s, the late-1960s run-up and the Great Inflation, and the post-pandemic surge. The moderate regime captures intermediate episodes, including the late 1950s, the post-disinflation period from the mid-1980s to the early 1990s, and 2005--07. The long sample therefore identifies regime-specific propagation from several distinct historical episodes rather than from the Great Inflation alone.

The threshold locations show that the regimes are not symmetric partitions of the historical inflation distribution. Annual inflation has a median of $2.07$ percent and a twenty-fifth percentile of $0.98$ percent over the estimation sample, while almost two-fifths of the pre-1950 observations are at or below zero. The lower and upper thresholds lie at approximately the $62$nd and $78$th percentiles, respectively. Thus, the estimated nonlinearity is concentrated in historically elevated inflation rather than reflecting a mechanical partition of the sample into equally populated states. The high-inflation regime covers roughly the upper fifth of the historical distribution, while the moderate regime occupies a comparatively narrow interval above its centre. \footnote{\citet{LopezSalidoLoria} show that the fit of quantile regressions to inflation is very similar to that obtained using a three-regime Markov-switching model, with the regimes providing a good approximation to key conditional quantiles.}

\subsection{\label{tailrisk}Macroeconomic Tail Risk}

Next, we use the estimated model to characterise state-dependent macroeconomic tail risk. Because the model jointly determines activity, inflation, financial conditions, and their volatilities, common shocks can alter the location, dispersion, and dependence of future outcomes. We use the 5th and 95th percentiles of the predictive distribution to measure downside and upside risks.

We construct recursive predictive distributions at horizons from one to eight quarters. At each forecast origin, the threshold and no-threshold models are estimated recursively using observations through that origin and simulated forward from the estimated state. Volatility and, in the threshold model, the inflation regime evolve endogenously.

Figure~\ref{fig:tailcompare} provides an illustrative comparison at the four-quarter horizon. It contrasts the percentile paths implied by the threshold model with those from a model that retains stochastic volatility but imposes a single propagation mechanism throughout the sample. The two specifications generate similar risk measures outside high-inflation periods. During high-inflation periods, however, the threshold model generally implies more adverse tail quantiles. The comparison therefore illustrates how allowing for state-dependent propagation changes the model-implied predictive distribution of activity and inflation. Because the specifications differ jointly in their coefficients, volatility dynamics, and level--volatility interactions, the comparison does not isolate a single channel.

\input{floats/fig_tailcompare}

The comparison is illustrative. We next evaluate the relative predictive performance of the threshold and no-threshold specifications over horizons from one to eight quarters. Table~\ref{tab:tailscores} reports two complementary diagnostics. Panel A reports RMSE ratios for the posterior predictive means and therefore assesses central forecast accuracy. Panel B reports outcome-weighted log-score differences in the sense of \citet{Amisano-Giacomini-07}, which place greater weight on realised outcomes far from the centre of the recursive estimation sample and hence assess relative density performance when outcomes are unusually distant from typical conditions.\footnote{The weight rises with the distance of the realised outcome from the centre of each recursive estimation sample. We report 100 times the average score difference. Outcome-weighted log scores are diagnostic rather than proper scoring rules \citep{gneiting-ranjan-11}.}

The forecast evidence is favourable to the threshold specification, but not uniformly so. Posterior-mean accuracy is broadly comparable across models: the threshold specification improves the credit spread at short horizons and inflation at longer horizons, whereas output-growth point forecasts are similar. The outcome-weighted log score favours the threshold specification in most variable--horizon comparisons, particularly at short horizons for GNP growth. The results therefore provide suggestive evidence that inflation-regime dependence can improve selected central forecasts and tail-outcome density diagnostics.

\input{tables/fms_table1}

Complementing the relative score comparisons, we assess the absolute calibration of each one-quarter-ahead predictive density using the probability integral transform (PIT), defined as the predictive cumulative distribution function evaluated at the realised outcome \citep{diebold-gunther-tay-98}. Under calibration, PITs are uniformly distributed. We test uniformity using the Kolmogorov--Smirnov (KS) and Cram\'er--von Mises (CvM) statistics of \citet{rossi-sekhposyan-19}.\footnote{We use the one-step-ahead critical values in \citet{rossi-sekhposyan-19}. As the models are recursively re-estimated, the test outcomes are indicative.} Unlike the outcome-weighted score, which compares models at realised outcomes, the PIT diagnostics assess each predictive density separately; the two exercises therefore need not rank the models identically.

\input{tables/tab_pit}

\input{floats/fig_pit}

Table~\ref{tab:pit} and Figure~\ref{fig:pit} show a relative calibration improvement for the threshold specification: both diagnostics are lower for every variable, especially inflation. Nevertheless, both models reject PIT uniformity for GNP growth and inflation, whereas the credit-spread densities are compatible with uniformity under either specification. The threshold model therefore improves calibration without delivering uniform gains across all forecast measures.

\FloatBarrier

\subsection{\label{leverage}State-Dependent Leverage and Volatility-in-Mean Feedback}

The model contains three related channels linking first and second moments: lagged macroeconomic outcomes affect subsequent volatility through $d_{j,m}$; level and volatility innovations are contemporaneously correlated through $\Sigma_{\eta e,(m)}$; and volatility feeds back into expected macroeconomic outcomes through the volatility-in-mean coefficients $b_{k,m}$. The threshold structure allows these channels, together with the remaining macroeconomic and volatility dynamics, to differ across inflation regimes.

\begin{figure}[!htbp]
\centering
\includegraphics[width=\textwidth]{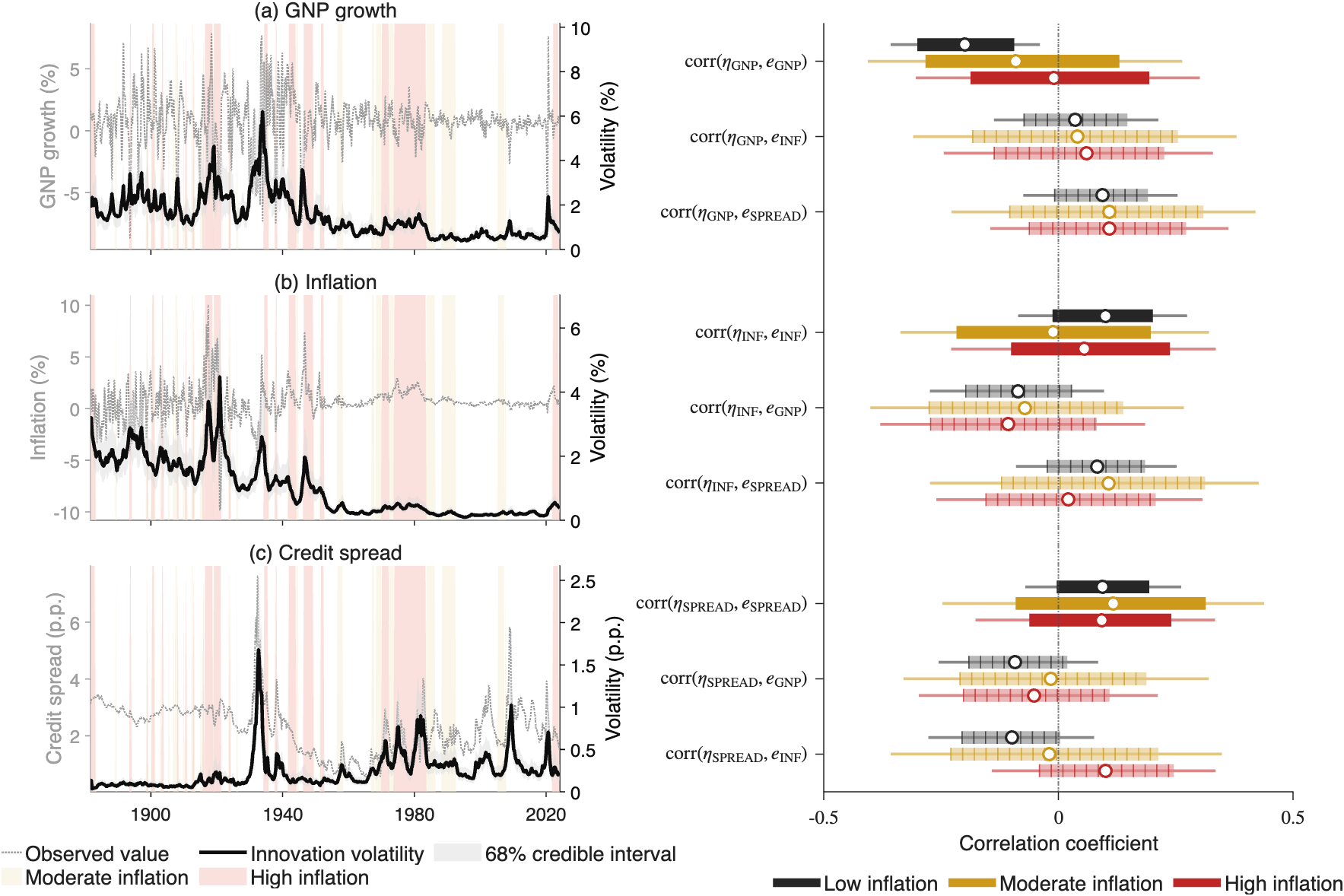}
\caption{\label{fig:volleverage}Innovation volatility and leverage correlations across inflation regimes.}
\caption*{\footnotesize \textbf{Notes:} \emph{Left:} dotted lines show observed values (left axis); solid lines show posterior median innovation standard deviations, $\exp(h_{i,t}/2)$, with 68\% credible intervals (right axis). Moderate- and high-inflation periods are shaded yellow and red, respectively. \emph{Right:} posterior medians with 68\%/90\% credible intervals for the regime-specific leverage block $\Sigma_{\eta e,(m)}$ of equation~\eqref{eq:sigmar}, whose $(i,j)$ element is $\operatorname{corr}(\eta_{i,t},e_{j,t})$, the correlation between the innovation to the volatility of variable $i$ and the level innovation of variable $j$; solid intervals denote own-variable correlations and lighter intervals cross-variable correlations. The vertical line denotes zero.}
\end{figure}

Figure~\ref{fig:volleverage} distinguishes two reduced-form objects. Its left-hand side plots each observed variable with the posterior median of its innovation volatility, $\exp(h_{i,t}/2)$, and the inflation-regime chronology. Appendix Table~\ref{tab:volobscorr} summarises this descriptive comovement over both the full historical sample and the postwar window used by \citet{CaldaraSVOL}. Its right-hand side reports the model's leverage estimates, the contemporaneous correlations $\operatorname{corr}(\eta_{i,t},e_{j,t})$ between first- and second-moment innovations.

Both panels report reduced-form features of the fitted model. The historical comovements on the left are descriptive, whereas the right-hand correlations summarise the contemporaneous dependence between level and volatility innovations. Neither identifies the response to an individual structural shock or a causal transmission channel. The figure also shows that inflation regimes are not simply volatility states: the Great Depression and the Global Financial Crisis lie predominantly in the low-inflation regime despite exceptionally high innovation volatility.

\citet{CaldaraSVOL} document three descriptive patterns: weak GDP growth coincides with higher GDP-growth innovation volatility; inflation and its innovation volatility comove positively on average, but inflation falls as its volatility rises during the Global Financial Crisis; and financial stress comoves with financial volatility. Appendix Table~\ref{tab:volobscorr} recovers the broad signs of these patterns in the shared postwar sample, but the longer historical sample shows that they vary across inflation environments.

Output provides the clearest connection between the descriptive evidence and the innovation-level estimates. Weak GNP growth tends to coincide with elevated output innovation volatility, a pattern evident in major recessions such as the Great Depression and the Global Financial Crisis, which are predominantly classified in the low-inflation regime. The innovation-level evidence is consistent with this pattern: in the low-inflation regime, negative output innovations co-occur with positive innovations to subsequent output volatility. This own-output leverage correlation attenuates in the moderate regime and is centred near zero when inflation is high. The estimates therefore suggest that, in the low-inflation state, adverse output news is associated with greater uncertainty about future output, whereas this reduced-form association weakens as inflation rises. This pattern is potentially relevant for downside growth risk, but does not identify the structural disturbance responsible for it.

For inflation, the positive postwar association with innovation volatility is concentrated in moderate- and high-inflation observations. It reverses in the low-inflation state and in the full historical sample. Own-inflation leverage is imprecisely estimated across regimes, so this descriptive evidence does not establish a stable relation between inflation innovations and innovations to subsequent inflation volatility.

Finally, the credit-spread evidence is the closest counterpart to the EBP result in \citet{CaldaraSVOL}: financial stress comoves positively with its innovation volatility in the shared postwar sample. In the full history, this relation is concentrated in the low-inflation state, which includes the Great Depression and the Global Financial Crisis, and does not extend uniformly to the other regimes. The posterior medians of own-spread leverage are positive but imprecisely estimated. Relatedly, the cross-variable leverage correlation between inflation innovations and innovations to subsequent credit-spread volatility changes sign across regimes, but is imprecisely estimated and should be interpreted only as suggestive evidence of a state-dependent reduced-form link between inflation and financial volatility. Given this posterior uncertainty, we do not interpret these reduced-form correlations as evidence about particular economic shocks. Section~\ref{SVAR} instead examines how identified shocks propagate through the full estimated system.

\subsection{Predictive risk of entering the high-inflation regime}

The model also provides a measure of near-term transition risk (regime-entry
risk) that complements inflation-at-risk. At each historical origin classified
outside the high-inflation regime, we use the full-sample posterior
and smoothed volatility states to compute the probability of entering
that regime at least once over the following four quarters.\footnote{These
probabilities are historical risk assessments conditional on
full-sample information, rather than forecasts constructed using
only the information available at each origin.}
Whereas inflation-at-risk describes the upper tail of the predictive
inflation distribution, the transition probability quantifies the
likelihood of moving into an environment in which the dynamics of
macroeconomic outcomes and volatility, and hence shock propagation,
differ.

\begin{figure}[!htbp]
\centering
\includegraphics[width=\textwidth]{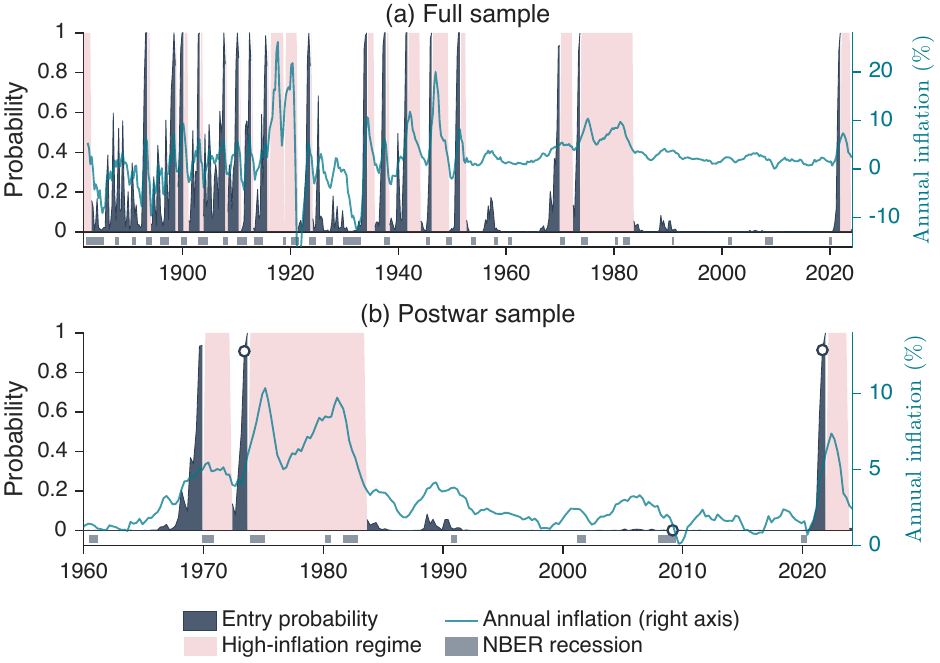}
\caption{\label{fig:regimerisk}Predictive probability of entering the high-inflation regime.}
\caption*{\footnotesize \textbf{Notes:} The filled area is the predictive probability that an origin outside the high-inflation regime enters it at least once over the following four quarters; the measure is undefined within that regime. The thin line is observed annual inflation (right axis), and high-inflation periods are shaded in red. Grey bars in the lane below the horizontal axis mark NBER recessions, from the peak to the trough quarter. Panel (a) plots the full sample and panel (b) the same series postwar. The inflation axis in panel (a) is truncated at $-16$ percent, so the 1921 deflation trough lies below the panel. Open circles mark 1973Q2, 2009Q1, and 2021Q3.}
\end{figure}

Figure~\ref{fig:regimerisk} shows that the probabilities rise ahead
of the highlighted transitions, indicating that the estimated model
associates the histories preceding these episodes with elevated
near-term transition risk. In both highlighted episodes, the
probability reaches $0.91$ two quarters before entry, compared with
a pre-1965 average of $0.25$. Most of the increase occurs in the
preceding quarter.

It is useful to note that this entry risk does not simply reflect rising volatility. It is essentially zero in 2009Q1 despite the exceptionally high innovation volatility documented in Figure~\ref{fig:volleverage}. The measure therefore distinguishes an extreme-volatility episode from one that carries a high near-term probability of entering a state with different propagation mechanisms.

Regime-entry risk and upside inflation risk are related but not redundant. Their positive, imperfect association (a correlation of $0.68$ at the four-quarter horizon) reflects their different objects. Inflation-at-risk is an intensive-margin measure of the severity of adverse inflation outcomes, whereas regime-entry risk is an extensive-margin measure of a transition into a state in which the propagation of shocks changes.

\section{\label{SVAR}Decomposing Macroeconomic Tail Risk}

The preceding analysis shows that the predictive distribution of macroeconomic outcomes varies across inflation regimes and provides reduced-form evidence of state dependence in some dimensions of mean--volatility interactions. We now turn to the structural drivers of these distributional changes. Specifically, we ask whether the shocks that account for movements in the centre of the predictive distribution are also the main drivers of growth- and inflation-at-risk, and whether their effects on tail risk depend on the inflation state and on the sign and size of the shock.

\subsection{Identification of shocks}

We identify four structural disturbances using the multi-shock max-share approach of \citet{Carriero02012025}: a \emph{business-cycle} shock targeting the forecast-error variance of GNP growth; a \emph{financial} shock targeting that of the credit spread; a \emph{macroeconomic-uncertainty} shock targeting the joint forecast-error variance of GNP-growth and inflation volatility; and a \emph{financial-uncertainty} shock targeting credit-spread volatility.

Candidate impact matrices take the form $A_0=C_0P_0$, where $C_0$ is the lower Cholesky factor of $\Omega_t$, the time-$t$ covariance matrix of the scaled reduced-form innovations, and $P_0$ is orthonormal. The first four columns of $A_0$ define the identified shocks. This rotation-based construction does not impose an economic recursive ordering. Because the model is nonlinear, with regime switching, volatility feedback, and time-varying shock scales, we compute simulated generalised impulse responses \citep{koop-pesaran-potter-96} and generalised forecast-error variance decompositions \citep{lanne-nyberg-16}, conditional on the initial state and cumulated over one year.

Following \citet{Carriero02012025}, we identify the four shocks jointly. The rotation $P_{0}$ maximises the equally weighted sum of the shares of the four targets explained by their corresponding shocks, subject to the requirement that each shock explains a larger share of its own target than of any other target.\footnote{These are the baseline inequality restrictions of \citet{Carriero02012025}, their equation 3.2. Cumulating the FEV shares over one year follows the variant in their Appendix~G.} We normalise the business-cycle shock to raise GNP growth on impact, the financial shock to raise the credit spread, and each uncertainty shock to raise its target volatility. Identification is repeated at every posterior draw and initial state. Thus, comparisons across inflation regimes hold the economic max-share criterion fixed, while allowing both the identified impact vector and its subsequent propagation to be state dependent.

\subsection{Nonlinear tail-risk responses}

For each posterior draw and initial state, we simulate the model with and without an impulse to an identified shock, using common future innovations in the two paths. The regime evolves endogenously along each simulated path. The response of growth-at-risk or inflation-at-risk is the difference between the relevant conditional quantiles of the shocked and baseline predictive distributions. We average these responses across histories within each inflation regime and report posterior medians and $68\%$ credible intervals. We use these responses to distinguish three forms of nonlinear propagation. State dependence refers to differences in the response to the same impulse across inflation regimes; sign asymmetry to departures from mirror symmetry between the responses to
equally sized positive and negative impulses; and size dependence to departures from proportionality as the magnitude of the impulse changes.

We focus on the low- and high-inflation regimes to provide a sharper contrast between inflation environments. The moderate regime remains fully embedded in the estimated model, entering the identification, simulated dynamics, and endogenous regime transitions. We do not report its responses separately because the smaller number of moderate-inflation histories results in less precise regime-specific estimates, while the two outer regimes provide the clearest comparison across inflation environments.

Figure~\ref{fig:risksize} plots the one-year-ahead response of the centre and tails to shocks ranging from $-5$ to $5$ standard deviations. Panel~(a) reports the predictive median, the $50^{th}$ percentile, as a centre-of-distribution benchmark; panel~(b) reports growth-at-risk (5th percentile) and inflation-at-risk (95th percentile). The predictive median is distinct from the conditional mean used in the conventional twelve-quarter impulse responses in the Appendix, since the predictive distribution can be asymmetric; it also serves as the centre-of-distribution benchmark in the decomposition below. \footnote{The Appendix also reports the twelve-quarter dynamics of risk following conventional one-standard-deviation shocks: uncertainty shocks have persistent effects on growth-at-risk at that scale, while state dependence is clearest for inflation-at-risk.} The larger impulses can be regarded as stress experiments designed to trace departures from proportional propagation, rather than representative shock realisations. Under proportional propagation, scaling an impulse scales its response proportionally, so responses trace a straight line through the origin. Curvature within either the positive or negative branch therefore indicates
size dependence, while departures from mirror symmetry between the responses
to equally sized positive and negative impulses indicate sign asymmetry. 

The predictive-median responses in panel~(a) are broadly proportional within each regime but there are some differences across regimes. Business-cycle shocks have somewhat stronger effects in the high-inflation regime. In contrast, financial shocks that increase (decrease) spreads decrease (increase) growth only in the low-inflation regime. The bottom panel shows that financial shocks instead have a stronger negative relationship with inflation during the high-inflation regime. Macroeconomic-uncertainty shocks (and financial-uncertainty shocks to a lesser extent) have a stronger effect on growth in the high-inflation regime. It is interesting to note that adverse uncertainty shocks are inflationary in this regime while their impact on inflation is close to zero when inflation is low.

Panel~(b) of the figure shows that larger departures from linearity within regimes are concentrated in the tails of the predictive distribution. Adverse uncertainty shocks lead to a larger deterioration of growth-at-risk than equally sized shocks that reduce uncertainty. These shocks also disproportionately increase the upside risk in inflation, especially during the high-inflation regime. Negative business-cycle shocks worsen left-tail risk in growth somewhat more than positive shocks. During the high-inflation regime, negative business-cycle and financial shocks are associated with a stronger increase in inflation upside risk.

\clearpage
\begin{landscape}
\thispagestyle{empty}
\vspace*{\fill}
\centering
\begin{minipage}[t]{0.48\linewidth}
\centering
\includegraphics[width=\linewidth]{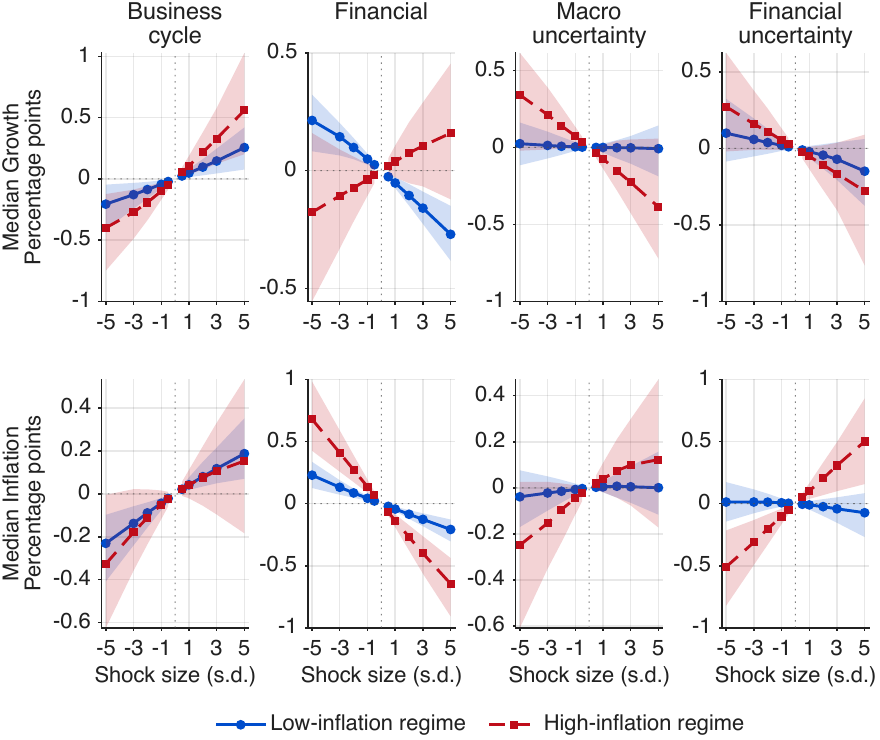}
\par\smallskip
\textbf{(a)} Predictive-median responses.
\end{minipage}\hfill
\begin{minipage}[t]{0.48\linewidth}
\centering
\includegraphics[width=\linewidth]{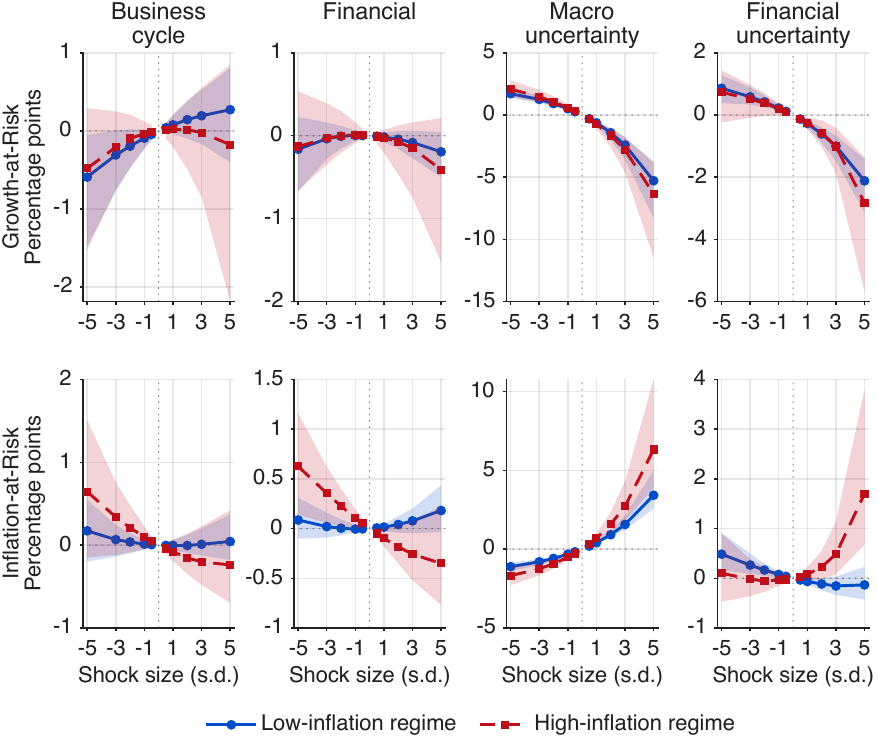}
\par\smallskip
\textbf{(b)} Tail-risk responses.
\end{minipage}

\begin{minipage}{\linewidth}
\captionof{figure}{Shock-size responses of the predictive distribution.}\label{fig:risksize}
\captionof*{figure}{\footnotesize \textbf{Notes:} Each panel reports one-year-ahead responses to impulses of $\pm 0.5$, $\pm 1$, $\pm 2$, $\pm 3$, and $\pm 5$ standard deviations. Panel~(a) reports the predictive median of GNP growth (top) and inflation (bottom). Panel~(b) reports growth-at-risk, the response of the $5^{th}$ percentile of GNP growth (top), and inflation-at-risk, the response of the $95^{th}$ percentile of inflation (bottom). Columns correspond to the business-cycle, financial, macroeconomic-uncertainty, and financial-uncertainty shocks. Blue solid lines average responses across low-inflation histories and red dashed lines across high-inflation histories; shaded areas are $68\%$ credible intervals across posterior draws. Under proportional propagation, the points in panel~(a) lie on a straight line through the origin.}
\end{minipage}
\vspace*{\fill}
\end{landscape}
\clearpage

\FloatBarrier
\subsection{Decomposing the drivers of tail risk}

The shocks that dominate conventional macroeconomic fluctuations need not be those that dominate movements in the tails. To quantify this distinction, we construct FEVD-style shares of the model-implied responses of growth- and inflation-at-risk and compare them with the corresponding decompositions of the predictive median. For each target quantile and shock size, the share assigned to an identified direction is its cumulated squared response divided by the corresponding sum over six orthogonal directions. Following the normalisation of \citet{lanne-nyberg-16}, we use these shares as an FEVD-style structural decomposition of model-implied tail-risk responses.

\begin{figure}[!htbp]
\centering
\includegraphics[width=\textwidth]{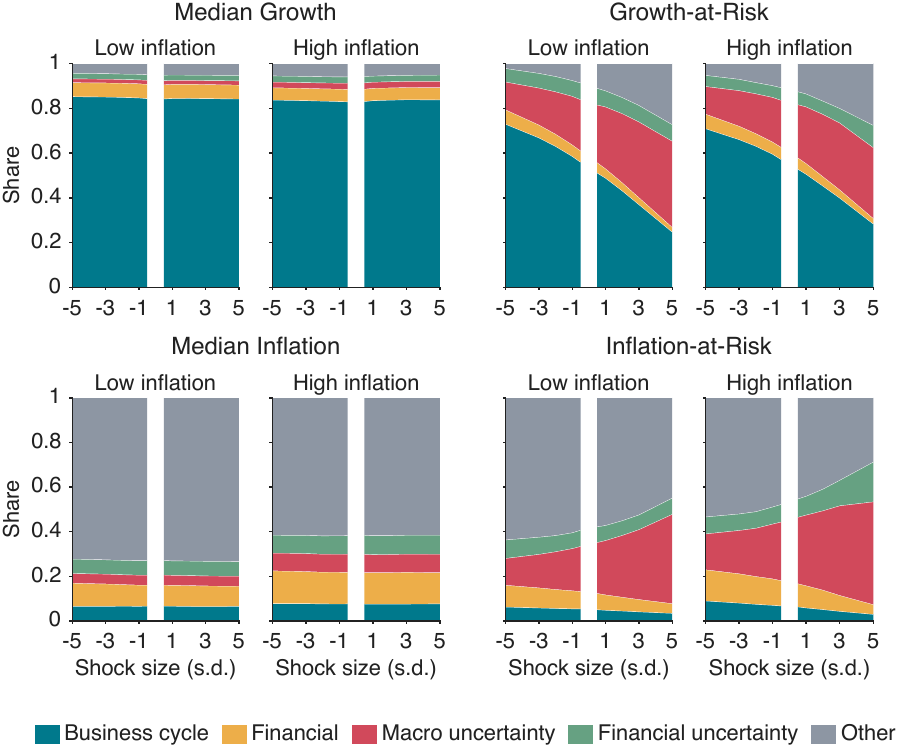}
\caption{\label{fig:qfevd_size}Structural decomposition of tail-risk responses against shock size.}
\caption*{\footnotesize \textbf{Notes:} Each panel reports a one-year decomposition based on cumulated squared responses to shocks ranging from $-5$ to $5$ standard deviations. A contribution is the squared response to an identified shock divided by the corresponding sum over six orthogonal directions. ``Other'' combines the two orthogonal directions not assigned an economic label by the max-share identification, together with any departure from exact additivity in the nonlinear construction. The top row concerns GNP growth and the bottom row inflation. The first two columns report the benchmark based on responses of the predictive median, the $50^{th}$ percentile; the final two report the tail-quantile shares, for the $5^{th}$ percentile of GNP growth and the $95^{th}$ percentile of inflation. Within each pair, the panels average over low- and high-inflation initial states respectively. For each posterior draw the five plotted shares sum to one by construction, since ``other'' is computed as one minus the sum of the four identified shares. Areas report posterior medians, across draws, of the individual shares; because these componentwise medians need not sum exactly to one, they are renormalised for display.}
\end{figure}

The first two columns of Figure~\ref{fig:qfevd_size} report shares based on responses of the predictive median, the centre-of-distribution benchmark; they decompose the predictive-median responses reported in panel~(a) of Figure~\ref{fig:risksize}. The final two columns report the analogous tail-quantile shares.

Figure~\ref{fig:qfevd_size} reveals a sharp distinction between the structural composition of the centre of the predictive distribution and that of tail risk. The business-cycle shock accounts for more than $80\%$ of the predictive-median decomposition of GNP growth, but its contribution to growth-at-risk is substantially smaller. Macroeconomic uncertainty fills part of this gap: despite its limited role at the median, it makes a material contribution to both growth- and inflation-at-risk. The sizeable ``other'' component for inflation-at-risk also shows that the four economically labelled shocks do not exhaust the structural directions relevant for inflation tail risk.

These differences are nonlinear and the tail shares vary across the negative and positive branches and with impulse size, whereas median-response shares are comparatively flat. The macroeconomic-uncertainty share of growth-at-risk rises with the size of an adverse macroeconomic-uncertainty impulse. In high-inflation histories, financial uncertainty likewise gains importance for inflation-at-risk following larger positive financial-uncertainty impulses.

The decomposition thus adds information beyond the response functions. It shows that the structural composition of tail risk depends on the state of the economy and on both the sign and size of the disturbance, even where the decomposition of the predictive median changes little. The threshold structure therefore implies state dependence not only in the predictive distribution of macroeconomic risk, but also in the structural composition of its model-implied responses.

\FloatBarrier
\section{\label{conclu}Conclusion}

This paper asks whether the shocks that account for conventional macroeconomic fluctuations also account for macroeconomic tail risk, and whether their propagation through the predictive distribution changes with the state of the economy. We address these questions with a threshold stochastic-volatility-in-mean VAR in which leverage, volatility-in-mean feedback, and the remaining macroeconomic and volatility dynamics vary across regimes. Predictive model selection supports a three-regime specification defined by inflation in a long U.S.\ sample.

The estimated regimes characterise distinct propagation environments rather than simply periods of high or low volatility. The Great Depression and the Global Financial Crisis belong predominantly to the low-inflation regime despite exceptionally high innovation volatility, while the reduced-form estimates suggest that mean--volatility dependence differs across inflation states.

Using max-share identification, the structural decomposition shows that the composition of tail risk differs from that of the predictive median. Business-cycle shocks dominate the median response of GNP growth but account for a substantially smaller share of growth-at-risk. Macroeconomic uncertainty, by contrast, has a limited role at the median yet makes a material contribution to both growth- and inflation-at-risk. Its contribution to growth-at-risk rises with the magnitude of a positive macroeconomic-uncertainty impulse. Financial uncertainty displays a corresponding pattern for inflation-at-risk in high-inflation histories.

The central implication is that a decomposition of the centre of the predictive distribution, or a measure of volatility alone, need not reveal the structural composition of macroeconomic tail risk. The shocks that dominate movements in the centre of the predictive distribution need not be those that dominate risks in its tails, and this distinction can itself depend on the state of the economy and on the sign and magnitude of the disturbance.

\bibliographystyle{ecta}
\bibliography{references/references}

\clearpage

\begin{center}
  {\large\bfseries Appendix}
\end{center}

\spacingset{1}

\appendix
\setcounter{figure}{0}\renewcommand{\thefigure}{A.\arabic{figure}}
\setcounter{table}{0}\renewcommand{\thetable}{A.\arabic{table}}
\setcounter{equation}{0}\renewcommand{\theequation}{A.\arabic{equation}}
\section{\label{app:model}Empirical model}

We consider the following regime-switching VAR model with stochastic volatility:

\begin{eqnarray}
\tilde{h}_{t+1} &=&\sum_{m=1}^{M}\left(\alpha_m +\theta_m \tilde{h}_{t}+%
\sum_{j=1}^{Q}d_{j,m}Y_{t-j}\right)\mathcal{I}(S_{t}=m)+\sum_{m=1}^{M}(\tilde{S}_{m}^{\frac{1}{2}})\mathcal{I}(S_{t}=m)\eta _{t}  \label{eq1-app} \\
Y_{t} &=&\sum_{m=1}^{M}\left(c_m+\sum_{j=1}^{P}\beta _{j,m}Y_{t-j}+\sum_{k=1}^{K}b_{k,m}\tilde{h}%
_{t-k}\right)\mathcal{I}(S_{t}=m)+H_{t}^{\frac{1}{2}}e_{t}  \label{eq2-app}
\end{eqnarray}

Here $\underset{N\times 1}{\underbrace{Y_{t}}}$ are a set
of endogenous variables. As explained below, $m=1,\dots,M$ denotes the regime with $\mathcal{I}(S_{t}=m)$ representing an indicator function that equals $1$ when regime $m$ is active.

The VAR system in equation~\eqref{eq2-app} is a regime switching VAR model with stochastic
volatility. The stochastic volatilities are denoted by
$\tilde{h}_{t}=[h_{1,t},h_{2,t},\dots,h_{N,t}]^{\prime }$ and $H_{t}=diag\left( \exp \left(
\tilde{h}_{t}\right) \right) $. As evident, $\tilde{h}_{t}$ is allowed to
have a lagged impact on the endogenous variables. The stochastic volatilities evolve
according to equation~\eqref{eq1-app}. The matrix $\tilde S_m=\operatorname{diag}(\tilde s_m)$,
with $\tilde s_m=[s_{1,m},s_{2,m},\ldots,s_{N,m}]$,
collects the marginal variances of the scaled volatility
innovations $\tilde S_m^{1/2}\eta_t$.
Their regime-$m$ covariance matrix is
$\tilde S_m^{1/2}\Sigma_{\eta,(m)}
(\tilde S_m^{1/2})^{\prime}$. Note that $%
\theta_m $ can be a full matrix with the elements of $\tilde{h}_{t}$ allowed
to have a dynamic relationship amongst themselves. The coefficients and error variances are subject to regime switching.

The disturbances $\varepsilon _{t}=\left(
\begin{array}{c}
\eta _{t} \\
e_{t}%
\end{array}%
\right) $, an $\left( 2N\times 1\right) $ vector, are distributed normally in each regime: $N(0,\Sigma_m )$ where the diagonal
elements of the $\left( 2N\times 2N\right) $ matrix $\Sigma_m $ are restricted to equal $1$:%
\begin{equation}
\Sigma_m =\left(
\begin{array}{cc}
\Sigma _{\eta,(m) } & \Sigma _{\eta e,(m)} \\
\Sigma _{\eta e,(m)}^{\prime } & \Sigma _{e,(m)}%
\end{array}%
\right)
\end{equation}%
where each block $\Sigma _{\eta ,(m)}$, $\Sigma _{e,(m)}$ and $\Sigma _{\eta e,(m)}$ is $\left( N\times N\right) $. In other words, the model allows for correlation between the shocks to the stochastic volatilities and the shocks to the level of the endogenous variables, with the strength of the correlation allowed to change across regimes; the $(i,j)$ element of $\Sigma _{\eta e,(m)}$ is $\mathrm{corr}(\eta _{i,t},e_{j,t})$. Because the same draw $\varepsilon _{t}=(\eta _{t}^{\prime },e_{t}^{\prime })^{\prime }$ enters $\tilde{h}_{t+1}$ (through $\eta _{t}$) and $Y_{t}$ (through $e_{t}$), the off-diagonal block $\Sigma _{\eta e,(m)}$ captures a same-period leverage correlation. Restricting the diagonal of $\Sigma_m $ to unity makes it a correlation matrix, so that all scale is carried by $\tilde{S}_{m}^{1/2}$ and $H_{t}^{1/2}$ in the covariance matrix below. The time-varying covariance matrix of the reduced form
residuals of the system in equations~\eqref{eq1-app} and~\eqref{eq2-app} can be written
as:
\begin{equation}
\Omega _{t}=\sum_{m=1}^{M}\mathcal{I}(S_{t}=m)\left(\left(
\begin{array}{cc}
\tilde{S}_{m}^{1/2} & 0 \\
0 & H_{t}^{1/2}%
\end{array}%
\right) \left(
\begin{array}{cc}
\Sigma _{\eta,(m) } & \Sigma _{\eta e,(m)} \\
\Sigma _{\eta e ,(m)}^{\prime } & \Sigma _{e,(m)}%
\end{array}%
\right) \left(
\begin{array}{cc}
\tilde{S}_{m}^{1/2} & 0 \\
0 & H_{t}^{1/2}%
\end{array}%
\right) ^{\prime }\right)
\label{covt}
\end{equation}

The regime switching process is defined by:
\begin{align}
S_{t}&=1\Longleftrightarrow Y_{i,t-d}\leq Y^{\ast,1 } \\
S_{t}&=m \Longleftrightarrow Y^{\ast,m } \ge  Y_{i,t-d}> Y^{\ast,m-1} \text{ for } m=2,\dots M-1   \\
S_{t}&=M \Longleftrightarrow  Y_{i,t-d}> Y^{\ast,M-1}
\end{align}%
where $Y_{i,t-d}$ is the $i$th endogenous variable lagged $d$ periods (i.e. the designated threshold variable, with $i$ a fixed pre-chosen index) and $Y^{\ast,m }$ for $m=1,\dots ,M-1$ are threshold values in an ascending order. Note that the threshold values and the delay parameter $d$ are treated as unknown parameters, with the latter allowed to take on
values $d=1,2,\dots,D$.
 For example, when the number of regimes $M=3$, there are $2$ threshold values and the regime switching process is governed by:
\begin{align}
 S_{t}&=1\Longleftrightarrow Y_{i,t-d}\leq Y^{\ast,1 } \\
S_{t}&=2 \Longleftrightarrow Y^{\ast,2 } \ge  Y_{i,t-d}> Y^{\ast,1 }  \\
S_{t}&=3 \Longleftrightarrow Y_{i,t-d}> Y^{\ast,2}
\end{align}
In the baseline specification used in the paper, $Y_t$ collects real GNP growth, inflation and a credit spread ($N=3$), and the preferred model sets $M=3$ with the threshold variable a transformation of inflation (Section~3 of the main text).

\subsection{Estimation}
We estimate the model using a Gibbs sampling algorithm. Several conditional posterior distributions are nonstandard, including those
of the thresholds, the regime-specific correlation matrices, and the
stochastic volatilities. In this section we describe the prior distributions and sketch the steps of the algorithm.

\subsection{Priors and starting values}

\paragraph{VAR coefficients}
Let
\[
\Gamma_m
=
\operatorname{vec}\!\left(
[\beta_{1,m},\ldots,\beta_{P,m},
 b_{1,m},\ldots,b_{K,m},c_m]^{\prime}
\right)
\]
collect the observation-equation coefficients in regime $m$.
The coefficient blocks are ordered to match the regressors
$Y_{t-1}^{\prime},\ldots,Y_{t-P}^{\prime},
\tilde h_{t-1}^{\prime},\ldots,\tilde h_{t-K}^{\prime},1$.
Thus, the intercept is the final regressor, and
$\operatorname{vec}$ stacks the columns of the transposed
coefficient matrix, collecting all coefficients of each equation
together.
Following \cite{Banbura-Giannone-Reichlin-10Paper}, we employ a
Normal prior for the coefficients in each regime $m$.
The priors are implemented by the dummy observations
$y_D$ and $x_D$:
\begin{equation}
\begin{gathered}
y_D=
\begin{bmatrix}
\tau^{-1}\operatorname{diag}(\gamma_1s_1,\ldots,\gamma_Ns_N)\\
0_{N(P-1)\times N}\\
0_{EX\times N}
\end{bmatrix},
\\[6pt]
x_D=
\begin{bmatrix}
\tau^{-1}\bigl[J_P\otimes\operatorname{diag}(s_1,\ldots,s_N)\bigr]
&0_{NP\times EX}\\
0_{EX\times NP}
&\operatorname{diag}(\tilde c_1^{-1},\ldots,\tilde c_{EX}^{-1})
\end{bmatrix},
\qquad J_P=\operatorname{diag}(1,\ldots,P).
\end{gathered}
\end{equation}
Here $\gamma_1,\ldots,\gamma_N$ are the prior means for the
own first-lag coefficients, obtained by estimating individual
AR(1) regressions, and $s_1,\ldots,s_N$ are the corresponding
residual standard deviations. The parameter $\tau$ controls
the overall prior tightness for the lagged endogenous variables,
while $\tilde c_\ell$ is the separate prior scale for the
$\ell$th additional regressor. For the observation equation,
$EX=NK+1$: the $NK$ lagged-volatility regressors are followed
by the intercept. We set $\tau=0.2$,
$\tilde c_\ell=1$ for $\ell=1,\ldots,NK$, and
$\tilde c_{EX}=1000$, giving an effectively flat prior for
the intercept. These dummy observations do not directly
implement a prior on the time-varying error covariance matrix.

The coefficient prior in each regime is
\[
\Gamma_m\sim N(\Gamma_0,P_0),
\]
where
\[
\Gamma_0
=
\operatorname{vec}\!\left[
(x_D^{\prime}x_D)^{-1}x_D^{\prime}y_D
\right],
\qquad
P_0=\bar S\otimes(x_D^{\prime}x_D)^{-1},
\]
and $\bar S=\operatorname{diag}(s_1^2,\ldots,s_N^2)$
contains the AR(1) residual variances obtained using the
training sample described below.

\paragraph{Initial log-volatility $\tilde{h}_{0}$}
Following \cite{cogley-sargent-05} we use a training sample of pre-sample
observations to set the prior for the initial state of the log-volatility
process. Let $\hat{v}^{ols}$ denote the OLS estimate of the VAR
covariance matrix estimated on the pre-sample data. Because $\tilde{h}_{t}$ is a
log-volatility ($H_{t}=diag(\exp (\tilde{h}_{t}))$), the prior on the initial value is
$\tilde{h}_{0}\sim N(\mu _{0},\underline{V}_{h})$ where $\mu _{0}=\log \left( diag\left(
\hat{v}^{ols}\right) \right) $ and $\underline{V}_{h}=0.1\,I$.

\paragraph{Elements of $\Sigma_m $}
The matrix $\Sigma_m$ is a correlation matrix: its diagonal elements are
restricted to equal $1$, so that all scale is carried by $\tilde{S}_{m}$ and
$H_{t}$ (see equation~\eqref{covt}). This is the ``separation strategy'' of
\cite{44b58bef-fe61-3589-95e9-7e6440f4511a}, which parameterises a covariance matrix in
terms of standard deviations and a correlation matrix. For the free elements
of $\Sigma_m$, the correlations $\rho_{kj,m}$ with $k>j$, we assume a flat
prior over the region where $\Sigma_m$ is positive definite. Unlike priors
placed on the elements of a triangular (Cholesky-type) factorisation of
$\Sigma_m$, this prior is invariant to the ordering of the variables in the
shock vector $\varepsilon_t$.

\paragraph{Parameters of equation \protect\eqref{eq1-app}}
The prior for the volatility-transition coefficient vector,
\[
\tilde\Gamma_m
=
\operatorname{vec}\!\left(
[\theta_m,d_{1,m},\ldots,d_{Q,m},\alpha_m]^{\prime}
\right),
\]
is constructed analogously to that for the observation-equation
coefficient vector $\Gamma_m$, with the regressors ordered as
$\tilde h_t^{\prime},Y_{t-1}^{\prime},\ldots,
Y_{t-Q}^{\prime},1$.
We assume an inverse-Gamma prior for the diagonal elements of $\tilde{S}_m$.

\paragraph{Threshold parameters}
We assume a normal prior for the threshold $Y^{\ast,m }$, $m=1,\dots,M-1$: $\mathcal{N}(\underline{Y^{\ast,m } },\underline{V_{Y^{\ast,m }}})$, where the prior mean is set as percentiles of the threshold variable, chosen to correspond to economically meaningful regime definitions (see the model-selection exercise in the main text). For example, in the three-regime case used in the paper, the mean for the first threshold is the $50^{th}$ percentile of the threshold variable and the mean for the second threshold is the $80^{th}$ percentile. $\underline{V_{Y^{\ast,m }}}$ is set tight, at $0.1$, so that each candidate specification is centred firmly on its intended regime definition; predictive performance in the main-text model-selection exercise then adjudicates between specifications. We assume a flat prior for the delay parameter.

\subsection{\label{app:gibbs}Gibbs algorithm}
The algorithm draws from the following conditional posterior distributions in each iteration.

\paragraph{Thresholds and delay parameter}
\begin{itemize}
    \item \textbf{Thresholds ($Y^{\ast}=\begin{pmatrix}
        Y^{\ast,1} \\
        Y^{\ast,2} \\
        \vdots \\
        Y^{\ast,M-1}
    \end{pmatrix}$)}: The thresholds are drawn one at a time, each conditional
    on the remaining threshold(s) and all other parameters, with the update
    order randomised in each iteration. The conditional posterior of the $m$th
    threshold is proportional to the product of the likelihood and the prior:
$$
F\left( Y^{\ast,m} \mid Y, \bm{\theta}_{-Y^{\ast,m}} \right) \propto f\left(Y \mid Y^{\ast}, \bm{\theta}_{-Y^{\ast}} \right) p\left(Y^{\ast} \right),
$$
where $Y$ represents the full set of observed data and $\bm{\theta}_{-Y^{\ast,m}}$ denotes all other model parameters held constant during this draw, \emph{including the remaining thresholds, the current volatility path $\tilde{h}_{1:T}$ and the regime-specific coefficients and covariances}. The prior restricts the thresholds to be in ascending order, $Y^{\ast,1} < \dots < Y^{\ast,M-1}$; we additionally require that each regime contain at least $n_{crit}$ observations. Candidate values that violate either restriction have zero posterior density. The threshold update uses the complete-data likelihood, with the
latent volatility states and the remaining parameters held fixed.
For a candidate regime allocation and $m=S_t$, define the scaled
stacked residual
\[
E_{t,m}=
\begin{pmatrix}
\tilde h_{t+1}-\alpha_m-\theta_m\tilde h_t
-\displaystyle\sum_{j=1}^{Q}d_{j,m}Y_{t-j}\\[6pt]
Y_t-c_m-\displaystyle\sum_{j=1}^{P}\beta_{j,m}Y_{t-j}
-\displaystyle\sum_{k=1}^{K}b_{k,m}\tilde h_{t-k}
\end{pmatrix}
=
G_{t,m}\varepsilon_{t,m},
\]
where
\[
G_{t,m}=
\begin{pmatrix}
\tilde S_m^{1/2} & 0\\
0 & H_t^{1/2}
\end{pmatrix},
\qquad
\varepsilon_{t,m}=
\begin{pmatrix}
\eta_t\\
e_t
\end{pmatrix}.
\]
The covariance of the standardised innovation vector is $\Sigma_m$,
whereas the covariance of the scaled residual vector is
$\Omega_t=G_{t,m}\Sigma_mG_{t,m}^{\prime}$, as in
equation~\eqref{covt}. The per-period Gaussian likelihood contribution
can equivalently be evaluated as
\[
\phi_{2N}(E_{t,m};0,\Omega_t)
=
\lvert\det G_{t,m}\rvert^{-1}
\phi_{2N}(\varepsilon_{t,m};0,\Sigma_m),
\]
where $\phi_{2N}(\,\cdot\,;0,V)$ denotes the $2N$-dimensional normal
density with mean zero and covariance $V$. Thus, evaluation using
standardised residuals must retain the scale Jacobian
$\lvert\det G_{t,m}\rvert^{-1}$. Multiplying the per-period contributions
gives the complete-data likelihood contribution to the threshold
update; no filtering is required in this step.

As a function of $Y^{\ast,m}$, the conditional posterior is a
\emph{step function}: the regime allocation, and hence the
likelihood, changes only when the threshold crosses an observed
value of the threshold variable $Y_{i,t-d}$, and is constant
between adjacent order statistics. A random-walk Metropolis step
on such a target requires careful tuning of the proposal scale
and traverses the flat segments slowly. We therefore draw each
threshold using the slice sampler of
\cite{feba456f-8373-3f4e-b93a-903e0caf3a0f}
(``shrinkage'' procedure), which requires only pointwise
evaluations of the conditional posterior, involves no tuning
parameters, and never rejects. One update of $Y^{\ast,m}$
proceeds as follows:

\begin{enumerate}
    \item Evaluate the log conditional posterior at the current value, $\ln F_0$, and draw the auxiliary slice level $\ln u = \ln F_0 + \ln v$ with $v \sim U(0,1)$.
    \item Initialise the bracket $[L, R]$ at the observed range of the threshold variable.
    \item Propose $z \sim U(L, R)$. If $\ln F(z) \geq \ln u$, accept $z$ as the new value of $Y^{\ast,m}$. Otherwise shrink the bracket towards the current value (set $L=z$ if $z$ lies below the current value, $R=z$ otherwise) and repeat step 3.
\end{enumerate}
Because the current value always lies within the slice, the shrinkage loop terminates with probability one; candidates that violate the ordering or $n_{crit}$ restrictions have zero posterior density and are shrunk away automatically. The update leaves the conditional posterior exactly invariant \citep{feba456f-8373-3f4e-b93a-903e0caf3a0f}, and in our applications requires roughly $15$ evaluations of $F(\cdot)$ per threshold per iteration.
    \item \textbf{Delay ($d$)}: \cite{Chen-98} shows that, under the flat prior on $d$, the conditional posterior for the delay parameter is a \textbf{multinomial distribution} with probabilities proportional to the likelihood:
$$
P(d \mid Y, \bm{\theta}_{-d}) \propto f\left( Y \mid d, \bm{\theta}_{-d} \right).
$$
The probability of drawing a specific delay $d \in \{1, \dots, D\}$ (where $D$ denotes the maximum delay allowed for)
is therefore given by:
$$
\frac{f\left( Y \mid d, \bm{\theta}_{-d} \right)}{\sum_{k=1}^{D}f\left( Y \mid d=k, \bm{\theta}_{-d} \right)},
$$
where $\bm{\theta}_{-d}$ denotes all model parameters other than $d$. For each candidate $d=k$ the threshold variable $Y_{i,t-k}$, and hence the implied regime allocation, is recomputed before the likelihood is evaluated.
\end{itemize}
After this step, the regime indicator $\mathbf{S}$ for all observations is updated based on the new thresholds and delay parameter, and the data are separated into $M$ sub-samples.

\paragraph{Regression coefficients}
This step jointly draws the coefficients of both the volatility (transition) equation~\eqref{eq1-app} and the observation equation~\eqref{eq2-app} in each regime; collect them as $\Pi_m=(\tilde{\Gamma}_m',\Gamma_m')'$.
Conditional on the regime indicator $S_t$, the parameters $\tilde{S}_m,H_{t}$ and $\Sigma_m $, the model can be written as a
SUR system with heteroscedasticity in each regime $m$%
\begin{eqnarray}
Z_{t,m} &=&X_{t,m}\Pi _{t}+E_{t,m}  \label{eq2a} \\
var\left( E_{t,m}\right)  &=&G_{t,m}\Sigma_m G_{t,m}^{\prime }  \notag
\end{eqnarray}%
where $G_{t,m}=diag\left( [\tilde{s}_m^{1/2},\exp \left( \tilde{h}_{t,m}\right)
^{1/2}]\right) $, so that $G_{t,m}\Sigma_m G_{t,m}^{\prime }$ is exactly the regime-$m$ time-varying covariance $\Omega_t$ of equation~\eqref{covt}, and
\begin{equation*}
Z_{t,m}=\left(
\begin{array}{c}
\tilde{h}_{1,t+1,m} \\
\tilde{h}_{2,t+1,m} \\
. \\
Y_{t,m}%
\end{array}%
\right) ,\quad X_{t,m}=\left(
\begin{array}{cccc}
x_{1t,m} & 0 & 0 & 0 \\
0 & x_{2t,m} & 0 & 0 \\
0 & 0 & . & 0 \\
0 & 0 & 0 & .%
\end{array}%
\right) ,\quad E_{t,m}=G_{t,m}\varepsilon_{t,m},\quad \varepsilon_{t,m}=\left(
\begin{array}{c}
\eta _{1t,m} \\
\eta _{2t,m} \\
. \\
e_{t,m}%
\end{array}%
\right)
\end{equation*}
Here $Z_{t,m}$ stacks the next-period log-volatilities $\tilde{h}_{t+1,m}$ (the transition block) and the current endogenous variables $Y_{t,m}$ (the observation block) in regime $m$, $\tilde{h}_{t,m}$ are the volatilities in regime $m$, $x_{it,m}$ denotes the observations of the regressors in the $i$th equation of the
system in regime $m$, and $E_{t,m}=G_{t,m}\varepsilon_{t,m}$ are the reduced-form residuals with covariance $\Omega_t$. The
coefficients in regime $m$ have a conditional posterior that is normal: $N\left( \Pi
_{T_m\backslash T_m},P_{T_m\backslash T_m}\right) $. Following \cite{carter-kohn-94} we
use the Kalman filter to compute the mean and variance of the conditional
posterior, accounting for the fact that the covariance matrix of the
residuals changes through time within regimes. To use the Kalman filter we define the
state transition as
\begin{equation*}
\Pi _{t}=\Pi _{t-1}.
\end{equation*}%
The Kalman filter is initialised at $\Pi _{0}$ and $P_{0|0}$, which are based
on the priors for the coefficients introduced above, and the recursions are
given by the following equations for $t=1,2,\dots,T_m$, i.e. the observations in regime $m$:

\begin{eqnarray*}
\Pi _{t\backslash t-1} &=&\Pi _{t-1\backslash t-1} \\
P_{t\backslash t-1} &=&P_{t-1\backslash t-1} \\
v _{t\backslash t-1} &=&Z_{t,m}-X_{t,m}\Pi _{t\backslash t-1} \\
f_{t\backslash t-1} &=&X_{t,m}P_{t\backslash t-1}X_{t,m}^{\prime }+\left(
G_{t,m}\Sigma_m G_{t,m}^{\prime }\right) \\
K_{t} &=&P_{t\backslash t-1}X_{t,m}^{\prime }f_{t\backslash t-1}^{-1} \\
\Pi _{t\backslash t} &=&\Pi _{t\backslash t-1}+K_{t}v _{t\backslash t-1}
\\
P_{t\backslash t} &=&P_{t\backslash t-1}-K_{t}X_{t,m}P_{t\backslash t-1}
\end{eqnarray*}%
Because the coefficients are constant within a regime (the state transition $\Pi_t=\Pi_{t-1}$ carries no innovation, so $P_{t\backslash t-1}=P_{t-1\backslash t-1}$), the final iteration of the Kalman filter at time $T_m$ already delivers the full-sample conditional posterior moments in regime $m$, $\Pi
_{T_m\backslash T_m}$ and $P_{T_m\backslash T_m}$; no backward-smoothing pass is required. The coefficient vector is then drawn as
\begin{equation*}
\Pi _{m}\sim N\left( \Pi _{T_m\backslash T_m},P_{T_m\backslash T_m}\right),
\end{equation*}%
with any draw that implies a nonstationary companion matrix discarded and redrawn.

\paragraph{Diagonal elements of $\tilde{S}_m$}
We use a Metropolis algorithm to draw the diagonal elements of
$\tilde S_m$ one at a time from their nonstandard conditional
posteriors. With the volatility states and remaining parameters held
fixed, we condition on the standardised observation innovations $e_t$.
Under the block convention used above, for $m=S_t$,
\begin{align*}
\mu_{\eta_t\mid e_t,m}
&\equiv
\operatorname{E}(\eta_t\mid e_t,S_t=m)
=
\Sigma_{\eta e,(m)}\Sigma_{e,(m)}^{-1}e_t,\\
\Sigma_{\eta_t\mid e_t,m}
&\equiv
\operatorname{Var}(\eta_t\mid e_t,S_t=m)
=
\Sigma_{\eta,(m)}
-
\Sigma_{\eta e,(m)}\Sigma_{e,(m)}^{-1}
\Sigma_{\eta e,(m)}^{\prime}.
\end{align*}
The log-volatility transition equation can therefore be written as
\begin{align}
\tilde h_{t+1}
-
\tilde S_m^{1/2}\mu_{\eta_t\mid e_t,m}
&=
\alpha_m+\theta_m\tilde h_t
+\sum_{j=1}^{Q}d_{j,m}Y_{t-j}
+\eta_{t,m}^{\ast},
\label{VARtrans-app}\\
\operatorname{Var}(\eta_{t,m}^{\ast}\mid e_t,S_t=m)
&=
\tilde S_m^{1/2}
\Sigma_{\eta_t\mid e_t,m}
\bigl(\tilde S_m^{1/2}\bigr)^{\prime}.
\notag
\end{align}
Here
$\eta_{t,m}^{\ast}
=\tilde S_m^{1/2}(\eta_t-\mu_{\eta_t\mid e_t,m})$
is the scaled conditional volatility innovation. With the states and
remaining parameters held fixed, it is uncorrelated with $e_t$.
The proposal density in each regime, $q\left( \cdot \right) $, is an
inverse Gamma:
\begin{equation*}
\tilde{S}_{m,\text{new}}\sim IG\left( v_{1,m},T_{1,m}\right) ,
\end{equation*}%
where the scale $v_{1,m}=\tilde{\eta}_{i,m}^{\prime }\tilde{\eta}_{i,m}+\underline{v}$ and the degrees of freedom $T_{1,m}=T_m+\underline{s}$, with $T_m$ the regime-$m$ sample size and $\tilde{\eta}_{i,m}$ the (marginal) residuals of the $i$th log-volatility equation~\eqref{eq1-app} in regime $m$.
The proposal is therefore built from the marginal transition residuals (a convenient approximation), while the acceptance step uses the true conditional likelihood implied by equation~\eqref{VARtrans-app}. The candidate is accepted with probability
\begin{equation*}
\alpha =\min\left( 1,\ \frac{g\left( \tilde{S}_{m,\text{new}}\mid Y,\bm{\theta}_{-\tilde{S}_m}\right) \,q\left( \tilde{S}_{m,\text{old}}\right) }{g\left(
\tilde{S}_{m,\text{old}}\mid Y,\bm{\theta}_{-\tilde{S}_m}\right) \,q\left( \tilde{S}_{m,\text{new}}\right) }\right),
\end{equation*}%
where $g\left( \cdot \right) \propto$ likelihood $\times$ prior is the target conditional posterior of $\tilde{S}_m$ given all other parameters at their values
drawn in previous steps, and $q\left( \cdot \right) $ is the inverse-Gamma proposal density evaluated at the old and new draws. Note that, because the proposal is an independence sampler, the proposal density is evaluated at the new draw in the denominator and at the old draw in the numerator. With the model in the form of a VAR (equation~\eqref{VARtrans-app}), the conditional likelihood can be evaluated easily.

\paragraph{Elements of $\Sigma_m $}
The draw is performed \emph{separately in each regime} $m$. Conditional on all
other parameters, the standardised stacked residuals
$\varepsilon_{t,m}=G_{t,m}^{-1}E_{t,m}=(\eta_{t,m}',e_{t,m}')'$ at the $T_m$
observations assigned to regime $m$ (i.e. $\{t:S_t=m\}$) are i.i.d.
$N(0,\Sigma_m)$. Under the flat prior on the correlations, the conditional
posterior of $\Sigma_m$ is therefore proportional to
\begin{equation}
g\left( \Sigma_m \mid Y, \bm{\theta}_{-\Sigma_m} \right) \propto
\left\vert \Sigma_m \right\vert^{-T_m/2}
\exp\left( -\tfrac{1}{2}\, tr\left( S_m \Sigma_m^{-1} \right) \right),
\qquad S_m=\sum_{t:S_t=m}\varepsilon_{t,m}\varepsilon_{t,m}^{\prime },
\label{sigmakernel}
\end{equation}
restricted to the set of positive-definite matrices with unit diagonal. There
is no standard distribution from which to sample on this restricted set. We
draw the free elements $\rho_{kj,m}$ one at a time (in random order within
each iteration), each from its conditional posterior given the remaining
correlations, using the same shrinkage slice sampler as for the thresholds.
\cite{44b58bef-fe61-3589-95e9-7e6440f4511a} show that, holding the other correlations
fixed, the set of values of $\rho_{kj,m}$ that keeps $\Sigma_m$ positive
definite is an interval. The slice update for $\rho_{kj,m}$ initialises the
bracket at $[-1,1]$; a candidate outside the positive-definite interval has
zero posterior density (the Cholesky factorisation of the implied $\Sigma_m$
fails) and is shrunk away automatically, so the feasible interval never needs
to be computed explicitly. Each evaluation of the kernel in equation
\eqref{sigmakernel} requires a single Cholesky factorisation of a
$2N \times 2N$ matrix, so the step is computationally trivial. As with the threshold draw, we use a shrinkage slice update of the full
conditional posterior, avoiding the need to tune a random-walk Metropolis
proposal. Candidate correlations outside the slice or the positive-definite
support are rejected, and the bracket is shrunk towards the current
correlation until an acceptable candidate is obtained. This procedure is repeated independently for each regime
$m=1,\dots,M$.

\paragraph{Elements of $H_t$}

Conditional on the observed data, the threshold parameters, and the
remaining model parameters, we update the latent volatility states
using an augmented state-space representation. The regime at each
date is determined by the observed threshold variable and the current
draw of the threshold parameters. We retain the timing of
equation~\eqref{eq1-app}: the innovation $\eta_t$ enters
$\tilde h_{t+1}$ and is contemporaneously correlated with the
observation innovation $e_t$.

For the empirical specification with $K=1$, define
\[
\digamma_t=
\begin{pmatrix}
\eta_{t+1}\\
\eta_t\\
\tilde h_{t+1}\\
\tilde h_t\\
\tilde h_{t-1}
\end{pmatrix},
\qquad
\digamma_{t-1}=
\begin{pmatrix}
\eta_t\\
\eta_{t-1}\\
\tilde h_t\\
\tilde h_{t-1}\\
\tilde h_{t-2}
\end{pmatrix}.
\]
Each block has dimension $N\times1$, so the augmented state has
dimension $5N\times1$, equal to $15\times1$ when $N=3$.

For $m=S_t$, its transition is
\[
\digamma_t=C_{t,m}+\Psi_m\digamma_{t-1}+N_t,
\]
where
\[
C_{t,m}=
\begin{pmatrix}
0_N\\
0_N\\
\alpha_m+\displaystyle\sum_{j=1}^{Q}d_{j,m}Y_{t-j}\\
0_N\\
0_N
\end{pmatrix},
\qquad
\Psi_m=
\begin{pmatrix}
0 & 0 & 0 & 0 & 0\\
I_N & 0 & 0 & 0 & 0\\
\tilde S_m^{1/2} & 0 & \theta_m & 0 & 0\\
0 & 0 & I_N & 0 & 0\\
0 & 0 & 0 & I_N & 0
\end{pmatrix},
\qquad
N_t=
\begin{pmatrix}
\eta_{t+1}\\
0_N\\
0_N\\
0_N\\
0_N
\end{pmatrix}.
\]
Here $0_N$ denotes an $N\times1$ zero vector, and every zero block
in $\Psi_m$ is $N\times N$. The third block row reproduces the
volatility equation,
\[
\tilde h_{t+1}
=
\alpha_m+\theta_m\tilde h_t
+\sum_{j=1}^{Q}d_{j,m}Y_{t-j}
+\tilde S_m^{1/2}\eta_t.
\]
The second block row carries forward $\eta_t$, while the fourth
and fifth block rows carry forward $\tilde h_t$ and
$\tilde h_{t-1}$, respectively.

The leading block introduces the next-period standardised innovation.
Under the model's innovation convention,
\[
\eta_{t+1}\mid S_{t+1}
\sim
N\!\left(0,\Sigma_{\eta,(S_{t+1})}\right).
\]
Consequently, the Gaussian innovation covariance associated with
this augmented transition is
\[
\tilde Q_t=
\begin{pmatrix}
\Sigma_{\eta,(S_{t+1})} & 0_{N\times4N}\\
0_{4N\times N} & 0_{4N\times4N}
\end{pmatrix}.
\]
The distinction between $S_t$ and $S_{t+1}$ reflects the indexing
of the augmented state: the deterministic transition produces
$\tilde h_{t+1}$ using the time-$t$ coefficients and $\eta_t$,
whereas the newly introduced innovation is dated $t+1$.
For the latent-state update, the required regimes can be computed
from the observed data and the threshold parameters. In particular,
$S_{T+1}$ depends only on observations available through $T$ because
the threshold delay is at least one period.

For $K>1$, append the additional volatility blocks
$\tilde h_{t-2},\ldots,\tilde h_{t-K}$ and extend the corresponding
identity-shift rows. The resulting state has dimension
$(K+4)N\times1$.

The observation-likelihood contribution at time $t$ uses the
current innovation $\eta_t$, the current volatility $\tilde h_t$,
and the lagged volatilities carried by the augmented state.
Retaining the covariance-block convention
\[
\Sigma_m=
\begin{pmatrix}
\Sigma_{\eta,(m)} & \Sigma_{\eta e,(m)}\\
\Sigma_{\eta e,(m)}^{\prime} & \Sigma_{e,(m)}
\end{pmatrix},
\]
the conditional mean and covariance of $e_t$, given $\eta_t$
and $S_t=m$, are
\begin{align*}
\mu_{e_t\mid\eta_t,m}
&=
\Sigma_{\eta e,(m)}^{\prime}
\Sigma_{\eta,(m)}^{-1}\eta_t,\\
\Sigma_{e_t\mid\eta_t,m}
&=
\Sigma_{e,(m)}
-
\Sigma_{\eta e,(m)}^{\prime}
\Sigma_{\eta,(m)}^{-1}
\Sigma_{\eta e,(m)}.
\end{align*}
The adjusted observation equation is therefore
\[
Y_t-H_t^{1/2}\mu_{e_t\mid\eta_t,m}
=
c_m+\sum_{j=1}^{P}\beta_{j,m}Y_{t-j}
+\sum_{k=1}^{K}b_{k,m}\tilde h_{t-k}
+\tilde e_t,
\]
where
\[
\tilde e_t
=
H_t^{1/2}\bigl(e_t-\mu_{e_t\mid\eta_t,m}\bigr).
\]
With the volatility states and remaining parameters held fixed,
the conditional covariance of this adjusted observation residual is
\[
V_{t,m}
=
H_t^{1/2}\Sigma_{e_t\mid\eta_t,m}
\bigl(H_t^{1/2}\bigr)^{\prime},
\qquad
H_t=\operatorname{diag}\!\left(\exp(\tilde h_t)\right).
\]
The $N\times N$ matrix $V_{t,m}$ is distinct from the
$2N\times2N$ covariance $\Omega_t$ of the full scaled innovation
vector in equation~\eqref{covt}.

Following recent developments in the seminal paper by \cite%
{RePEc:bla:jorssb:v:72:y:2010:i:3:p:269-342}, we employ a particle Gibbs
step to sample from the conditional posterior of $\digamma _{t}$. \cite%
{RePEc:bla:jorssb:v:72:y:2010:i:3:p:269-342} show how a version of the
particle filter, conditioned on a fixed trajectory for one of the particles
can be used to produce draws that result in a Markov Kernel with a target
distribution that is invariant. However, the usual problem of path
degeneracy in the particle filter can result in poor mixing in the original
version of particle Gibbs. Recent developments suggest that small
modifications of this algorithm can largely alleviate this problem. In
particular, \cite{JMLR:v15:lindsten14a} propose the addition of a step that
involves sampling the `ancestors' or indices associated with the particle
that is being conditioned on. They show that this results in a substantial
improvement in the mixing of the algorithm even with a few particles. As
explained in \cite{JMLR:v15:lindsten14a}, ancestor sampling breaks the
reference path into pieces and this causes the particle system to collapse
towards something different than the reference path. In the absence of this
step, the particle system tends to collapse to the conditioning path. We
employ particle Gibbs with ancestor sampling in this step. The deterministic augmented-state transition and the observation likelihood at time $t$ use the parameters of regime $S_t$. The
Gaussian covariance for the leading innovation $\eta_{t+1}$ uses regime $S_{t+1}$. The transition and observation factors entering
ancestor sampling must preserve this same indexing.

Let $\digamma _{t}^{\left( i-1\right) }$ denote the fixed
trajectory, for $t=1,2,\ldots,T$ obtained in the previous draw of the Gibbs
algorithm. We denote all the parameters of the model by $\Xi $, and $j=1,2,\ldots,%
\tilde{M}$ indexes the particles.

For particle $j$ at time $t$, let $\tilde h_{t,[0]}^{(j)}$
denote the current log-volatility vector,
$\tilde h_{t,[-k]}^{(j)}$ its $k$th lag carried by that particle,
and $\eta_t^{(j)}$ its current standardised volatility innovation.
For $m=S_t$, define
\begin{align*}
H_t^{(j)}
&=
\operatorname{diag}
\left(\exp\bigl(\tilde h_{t,[0]}^{(j)}\bigr)\right),\\
\mu_{e_t\mid\eta_t,m}^{(j)}
&=
\Sigma_{\eta e,(m)}^{\prime}
\Sigma_{\eta,(m)}^{-1}\eta_t^{(j)},\\
V_{t,m}^{(j)}
&=
\bigl(H_t^{(j)}\bigr)^{1/2}
\Sigma_{e_t\mid\eta_t,m}
\left[\bigl(H_t^{(j)}\bigr)^{1/2}\right]^{\prime}.
\end{align*}
The corresponding adjusted observation residual is the $N\times1$
column vector
\[
u_t^{(j)}
=
Y_t-c_m-\sum_{\ell=1}^{P}\beta_{\ell,m}Y_{t-\ell}
-\sum_{k=1}^{K}b_{k,m}\tilde h_{t,[-k]}^{(j)}
-\bigl(H_t^{(j)}\bigr)^{1/2}
\mu_{e_t\mid\eta_t,m}^{(j)}.
\]
Omitting the normalising constant common to all particles, the
observation-likelihood weight and its normalisation are
\begin{align*}
w_t^{(j)}
&=
\left|V_{t,m}^{(j)}\right|^{-1/2}
\exp\left\{
-\frac{1}{2}
\bigl(u_t^{(j)}\bigr)^{\prime}
\bigl(V_{t,m}^{(j)}\bigr)^{-1}
u_t^{(j)}
\right\},\\
p_t^{(j)}
&=
\frac{w_t^{(j)}}{\displaystyle\sum_{a=1}^{\tilde M}w_t^{(a)}}.
\end{align*}

The conditional particle filter with
ancestor sampling proceeds in the following steps:

\begin{enumerate}
\item For $t=1$

\begin{enumerate}
\item Draw $\digamma _{1}^{(j)}\backslash \digamma _{0}^{(j)}$,$\Xi $ for $%
j=1,2,\ldots,\tilde{M}-1$. Fix $\digamma _{1}^{(\tilde{M})}=\digamma _{1}^{\left(
i-1\right) }$

\item Compute the observation-likelihood weights $w_1^{(j)}$
and their normalised values $p_1^{(j)}$ using the expressions
above with $m=S_1$, for $j=1,\dots,\tilde M$.
\end{enumerate}

\item For $t=2$ to $T$ (with active regime $m=S_t$ at each $t$)

\begin{enumerate}
\item For each nonreference particle $j=1,\ldots,\tilde M-1$,
draw an ancestor index $a_t^{(j)}\in\{1,\ldots,\tilde M\}$ with
\[
\Pr\!\left(a_t^{(j)}=a\right)=p_{t-1}^{(a)},
\qquad a=1,\ldots,\tilde M.
\]
Use $\digamma_{t-1}^{(a_t^{(j)})}$ as its resampled ancestor.

\item Propagate each nonreference particle
$j=1,\dots,\tilde M-1$ from its resampled ancestor:
\[
\digamma_t^{(j)}
=
C_{t,S_t}
+
\Psi_{S_t}\digamma_{t-1}^{(a_t^{(j)})}
+
N_t^{(j)},
\]
where $N_t^{(j)}$ has nonzero entries only in its leading
$N$-dimensional block. That block is drawn from
$N(0,\Sigma_{\eta,(S_{t+1})})$, in accordance with the
augmented-state transition above.

\item Fix $\digamma _{t}^{(\tilde{M})}=\digamma _{t}^{\left( i-1\right) }$

\item Sample $a_{t}^{(\tilde{M})}$ with probability defined by weights:
\begin{equation}
w_{t-1}^{(j)}\prod_{s=t}^{\min\{T,t-1+\mathcal{L}\}}g(Y_{s}|\digamma^{(j)}_{1:t-1},\digamma^{(i-1)}_{t:s})f(\digamma^{(i-1)}_{s}|\digamma^{(j)}_{1:t-1},\digamma^{(i-1)}_{t:s-1})
\end{equation}
At each date $s$ in the ancestor-sampling window, the observation factor $g(\cdot)$ uses regime $S_s$. The augmented transition factor
$f(\cdot)$ uses the deterministic coefficients of regime $S_s$ and the Gaussian covariance
$\Sigma_{\eta,(S_{s+1})}$ for its leading innovation $\eta_{s+1}$. As discussed in \cite{JMLR:v15:lindsten14a}, this formula is an approximation that can be used when, as in our case, the transition density is degenerate.

\item Update the observation-likelihood weights $w_t^{(j)}$
and their normalised values $p_t^{(j)}$ using the expressions
above with $m=S_t$, for $j=1,\dots,\tilde M$.
\end{enumerate}

\item End

\item Sample $\digamma _{t}^{\left( i\right) }$ with $\Pr \left( \digamma
_{t}^{\left( i\right) }=\digamma _{t}^{\left( j\right) }\right) \propto $ $%
p_{T}^{(j)}$ to obtain a draw from the conditional posterior distribution.
\end{enumerate}
We use $\tilde{M}=20$ particles in our application. The initial values $\mu
_{0}$ defined above are used to initialise step 1 of the filter.

\section{\label{app:sim}Simulation Design and Calibration}

To assess finite-sample recovery, we conduct a simulation exercise in which data are generated from the state-space system described in Appendix~\ref{app:model}, in two controlled experiments: a two-regime model ($M=2$, one threshold) and a three-regime model ($M=3$, two thresholds).

\paragraph{Model dimensions and sample size.}
In both experiments the data-generating process (DGP) has $N=2$ endogenous variables, one lag in the observation equation ($P=1$), one lag of the volatility in the conditional mean ($K=1$), and one lag of the endogenous variables in the volatility equation ($Q=1$). We simulate $700$ observations and discard the first $100$ as burn-in, leaving $600$ observations.

\paragraph{Threshold and regime configuration.}
The threshold variable is the second endogenous variable ($i=2$) with delay $d=1$. In the two-regime DGP the threshold is $Y^{\ast}=-0.6$. In the three-regime DGP the thresholds are $Y^{\ast,1}=-0.9$ and $Y^{\ast,2}=0.04$, chosen close to the $33^{rd}$ and $67^{th}$ percentiles of the marginal distribution of the threshold variable so that each regime holds roughly one third of the sample.

\paragraph{Parameter values.}
The remaining true parameter values are reported in Table~\ref{tab:dgp}; the error correlation matrices are
\begin{gather*}
\Sigma_1 =\begin{pmatrix} 1 & 0.2 & 0.3 & -0.4 \\ 0.2 & 1 & 0.6 & 0.2 \\
0.3 & 0.6 & 1 & -0.2 \\ -0.4 & 0.2 & -0.2 & 1 \end{pmatrix},\qquad
\Sigma_2 =\begin{pmatrix} 1 & -0.3 & 0.1 & 0.5 \\ -0.3 & 1 & -0.5 & 0.1 \\
0.1 & -0.5 & 1 & 0.3 \\ 0.5 & 0.1 & 0.3 & 1 \end{pmatrix}, \\[1ex]
\Sigma_3 =\begin{pmatrix} 1 & 0.1 & -0.2 & 0.3 \\ 0.1 & 1 & 0.4 & -0.3 \\
-0.2 & 0.4 & 1 & 0.2 \\ 0.3 & -0.3 & 0.2 & 1 \end{pmatrix},
\end{gather*}
with the shock ordering $\varepsilon_t=(\eta_{1t},\eta_{2t},e_{1t},e_{2t})'$,
so that the off-diagonal blocks capture the leverage correlations between the
volatility and observation shocks. Regimes 1 and 2 of the three-regime DGP share the parameter
values of the two-regime DGP (Table~\ref{tab:dgp}).

\begin{table}[htbp]
\centering
\caption{True parameter values in the simulation DGPs}
\label{tab:dgp}
\footnotesize
\begin{tabular}{lccc}
\toprule
 & Regime 1 & Regime 2 & Regime 3 (three-regime DGP) \\
\midrule
$\theta_m$ & $\begin{pmatrix} 0.85 & -0.10 \\ 0.10 & 0.85 \end{pmatrix}$ &
$\begin{pmatrix} 0.75 & -0.20 \\ 0.10 & 0.75 \end{pmatrix}$ &
$\begin{pmatrix} 0.65 & -0.30 \\ 0.10 & 0.65 \end{pmatrix}$ \\[2ex]
$d_{1,m}$ & $\begin{pmatrix} -0.05 & 0.01 \\ -0.05 & 0.01 \end{pmatrix}$ &
$\begin{pmatrix} -0.10 & 0.01 \\ -0.05 & 0.01 \end{pmatrix}$ &
$\begin{pmatrix} -0.15 & 0.01 \\ -0.10 & 0.01 \end{pmatrix}$ \\[2ex]
$\alpha_m$ & $\begin{pmatrix} 0 \\ 0 \end{pmatrix}$ &
$\begin{pmatrix} -0.5 \\ 0 \end{pmatrix}$ &
$\begin{pmatrix} 0.3 \\ 0 \end{pmatrix}$ \\[2ex]
$diag(\tilde{S}_m)$ & $(0.8,\ 0.8)$ & $(1,\ 1)$ & $(1.2,\ 1.2)$ \\[1ex]
$\beta_{1,m}$ & $\begin{pmatrix} 0.5 & -0.1 \\ 0.1 & 0.5 \end{pmatrix}$ &
$\begin{pmatrix} 0.5 & -0.1 \\ 0.1 & 0.5 \end{pmatrix}$ &
$\begin{pmatrix} 0.5 & -0.1 \\ 0.1 & 0.5 \end{pmatrix}$ \\[2ex]
$b_{1,m}$ (two-regime) & $\begin{pmatrix} -0.05 & 0.01 \\ -0.05 & 0.01 \end{pmatrix}$ &
$\begin{pmatrix} -0.10 & 0.01 \\ -0.10 & 0.01 \end{pmatrix}$ & --- \\[2ex]
$b_{1,m}$ (three-regime) & $\begin{pmatrix} -0.05 & 0.01 \\ -0.05 & 0.01 \end{pmatrix}$ &
$\begin{pmatrix} -0.10 & 0.01 \\ -0.10 & 0.01 \end{pmatrix}$ &
$\begin{pmatrix} -0.15 & 0.01 \\ -0.15 & 0.01 \end{pmatrix}$ \\[2ex]
$c_m$ & $(0.3,\ -0.3)'$ & $(-0.3,\ -0.3)'$ & $(0.6,\ -0.3)'$ \\
\bottomrule
\end{tabular}
\end{table}

\paragraph{Sampler settings.}
The models are estimated with the algorithm described in Appendix~\ref{app:model}, using the true
lag structure, a maximum delay of $D=2$, a minimum regime size of $10\%$ of
the sample and $\tilde{M}=20$ particles in the particle Gibbs step. The
sampler is run for $5{,}000$ iterations with a burn-in of $1{,}000$ in the
two-regime case, and for $12{,}000$ iterations with a burn-in of $7{,}000$ in
the three-regime case, retaining every second draw. The longer burn-in in the
three-regime case reflects the richer threshold configuration space of that
model.

\begin{figure}[htbp]
\centering
\includegraphics[width=\textwidth]{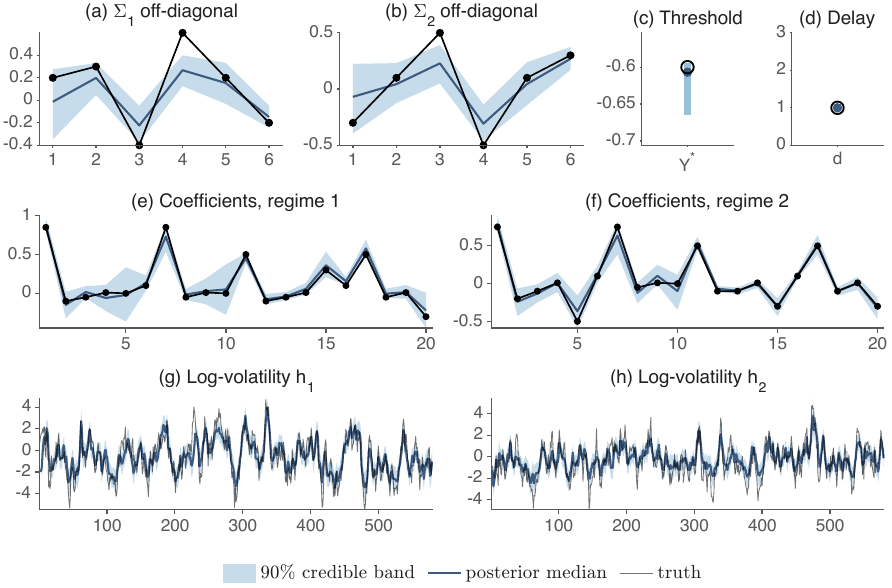}
\caption{Simulation experiment, two-regime model: posterior estimates versus
true values. Shaded areas are the $5^{th}$--$95^{th}$ posterior percentile
bands and the solid blue lines the posterior medians; for the log-volatilities
in panels (g) and (h) the band is the $16^{th}$--$84^{th}$ percentile range.
Black lines and markers denote the true values. Panels (a) and (b) show the
six free elements of each regime's error correlation matrix, indexed along the
horizontal axis; panels (e) and (f) show the stacked coefficients of the
volatility and observation equations; panels (g) and (h) show the
log-volatility paths over the simulated sample. Panels (c) and (d) report the
threshold and the delay parameter: the vertical bar is the
$5^{th}$--$95^{th}$ percentile range, the filled marker the posterior median
and the open circle the true value.}
\label{fig:sim2}
\end{figure}

\begin{figure}[htbp]
\centering
\includegraphics[width=\textwidth]{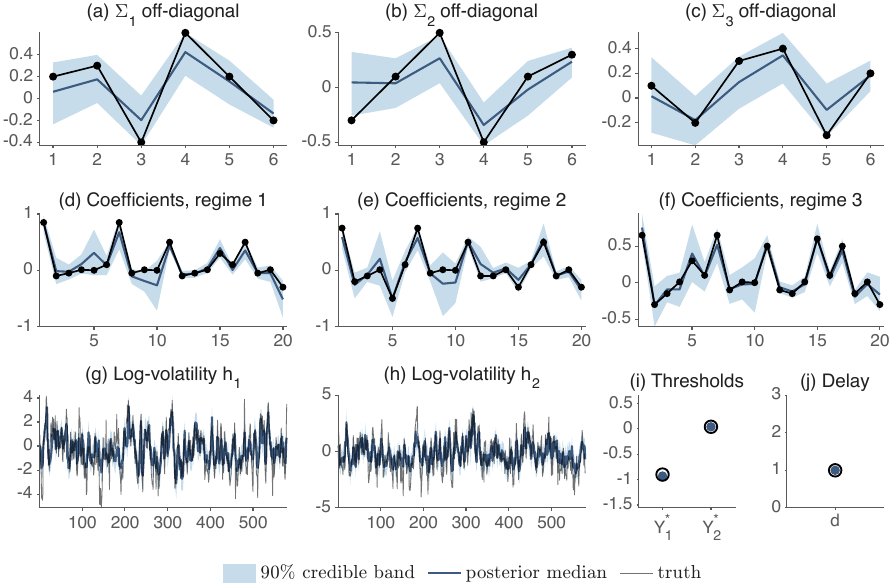}
\caption{Simulation experiment, three-regime model: posterior estimates versus
true values. Panels (a)--(c) show the six free elements of each regime's error
correlation matrix; panels (d)--(f) the stacked coefficients of the volatility
and observation equations; panels (g) and (h) the log-volatility paths over
the simulated sample; panels (i) and (j) the two thresholds and the delay
parameter. Bands, lines and markers are as in Figure~\ref{fig:sim2}.}
\label{fig:sim3}
\end{figure}

Figures~\ref{fig:sim2} and~\ref{fig:sim3} compare posterior
medians with the true values. Bands show the 5th--95th
posterior percentiles, except in the log-volatility panels,
where they show the 16th--84th percentiles. Overall, the algorithm performs reasonably well. The thresholds and the delay are estimated precisely in both
experiments: the posterior of the threshold(s) is tightly concentrated around
the true value(s) and the posterior probability of the true delay equals one.
 The estimated
log-volatility paths track the true paths closely, with the true volatilities
inside the posterior bands almost everywhere, and the credible intervals for
the regression coefficients of both equations cover the true values. The posterior medians of the error correlations in $\Sigma_m$ generally have the same sign as their true values but are somewhat attenuated towards zero for some of the larger correlations, reflecting the fact that the data are only weakly informative about some correlations involving latent volatility shocks in sub-samples.

\clearpage

\section{\label{app:data}Data}

The raw data span 1875Q2--2024Q1 and contain three quarterly U.S.\ variables: real GNP growth, inflation measured by the GNP deflator, and a credit spread capturing financial conditions. After transformations, lags, and the pre-sample used to calibrate the priors, the estimation sample begins in 1881Q3. The long historical series are constructed by combining pre-1947 data from the NBER tables accompanying \emph{The American Business Cycle} with more recent data from FRED and the Global Financial Database (GFD). Table~\ref{tab:data} lists the constituent series and the point at which each variable is spliced from the historical to the modern source.

\begin{table}[htbp]
\centering
\caption{Construction of the estimation sample}
\label{tab:data}
\footnotesize
\begin{tabular}{lll}
\toprule
Variable & Pre-1947 source & 1947-- source \\
\midrule
Real GNP growth & NBER RGNP72 & FRED real GNP (GNPC1) \\
Inflation & NBER GNP deflator (GNPD72) & FRED GNP deflator (GNPDEF) \\
Corporate bond yield & NBER CORPYIELD & FRED Moody's Baa yield (BAA) \\
Government bond yield & \multicolumn{2}{l}{GFD 10-year constant-maturity yield (IGUSA10d), full sample} \\
\bottomrule
\end{tabular}
\vspace{0.4em}

\begin{minipage}{0.85\textwidth}
\footnotesize
\textbf{Notes:} The credit spread used in estimation is the difference between the corporate bond yield and the 10-year government bond yield. Real GNP growth and inflation are constructed as log differences of the corresponding level series before and after splicing.
\end{minipage}
\end{table}

\section{\label{addres}Additional results}

\subsection{Descriptive mean--volatility comovement}

Table~\ref{tab:volobscorr} reports the descriptive correlations between each observed variable
and its estimated innovation volatility in the full historical sample and in
the postwar window considered by \citet{CaldaraSVOL}. These correlations
summarise reduced-form comovement and are distinct from the model's leverage
parameters.

\begingroup
\begin{table}[htbp]
\centering
\footnotesize
\caption{\label{tab:volobscorr}Descriptive correlations between observed values and innovation volatility}
\setlength{\tabcolsep}{8pt}
\renewcommand{\arraystretch}{1.10}
\begin{tabular}{@{}lcccc@{}}
\toprule
 & & \multicolumn{3}{c}{\textbf{Inflation regime}} \\
\cmidrule(l){3-5}
\rowcolor{gray!10}\textbf{Sample} & \textbf{All quarters} & \textbf{Low} & \textbf{Moderate} & \textbf{High} \\
\midrule
\rowcolor{gray!10}\multicolumn{5}{@{}l}{\textbf{GNP growth}} \\
Full sample & $-0.12$ & $-0.08$ & $-0.19$ & $-0.17$ \\
1954Q2--2019Q4 & $-0.12$ & $-0.06$ & $-0.17$ & $-0.02$ \\
\midrule
\rowcolor{gray!10}\multicolumn{5}{@{}l}{\textbf{Inflation}} \\
Full sample & $-0.08$ & $-0.16$ & $-0.06$ & $-0.07$ \\
1954Q2--2019Q4 & $0.70$ & $-0.20$ & $0.13$ & $0.73$ \\
\midrule
\rowcolor{gray!10}\multicolumn{5}{@{}l}{\textbf{Credit spread}} \\
Full sample & $0.33$ & $0.50$ & $-0.48$ & $0.04$ \\
1954Q2--2019Q4 & $0.62$ & $0.78$ & $0.47$ & $0.71$ \\
\bottomrule
\end{tabular}
\caption*{\footnotesize \textit{Notes:} Correlation between each observed variable $Y_{i,t}$ and its posterior median innovation standard deviation $\exp(h_{i,t}/2)$, for all quarters and for the quarters assigned to each modal inflation regime. The full sample is the model's estimation sample, 1881Q3--2024Q1 (571 quarters: 352 low-, 94 moderate- and 125 high-inflation). It begins 25 quarters after the start of the data, 1875Q2: five quarters are absorbed by the two lags and by the four-quarter threshold variable lagged one quarter, and the first 20 quarters serve as a pre-sample for calibrating the priors. 1954Q2--2019Q4 is the estimation sample of \citet{CaldaraSVOL} (263 quarters: 155, 60 and 48). Volatilities and regimes are from the full-sample estimates. These are descriptive statistics, not the model's leverage parameters.}
\end{table}
\endgroup
\FloatBarrier

\subsection{Conventional and one-standard-deviation structural responses}

\begin{figure}[!htbp]
\centering
\includegraphics[width=\textwidth]{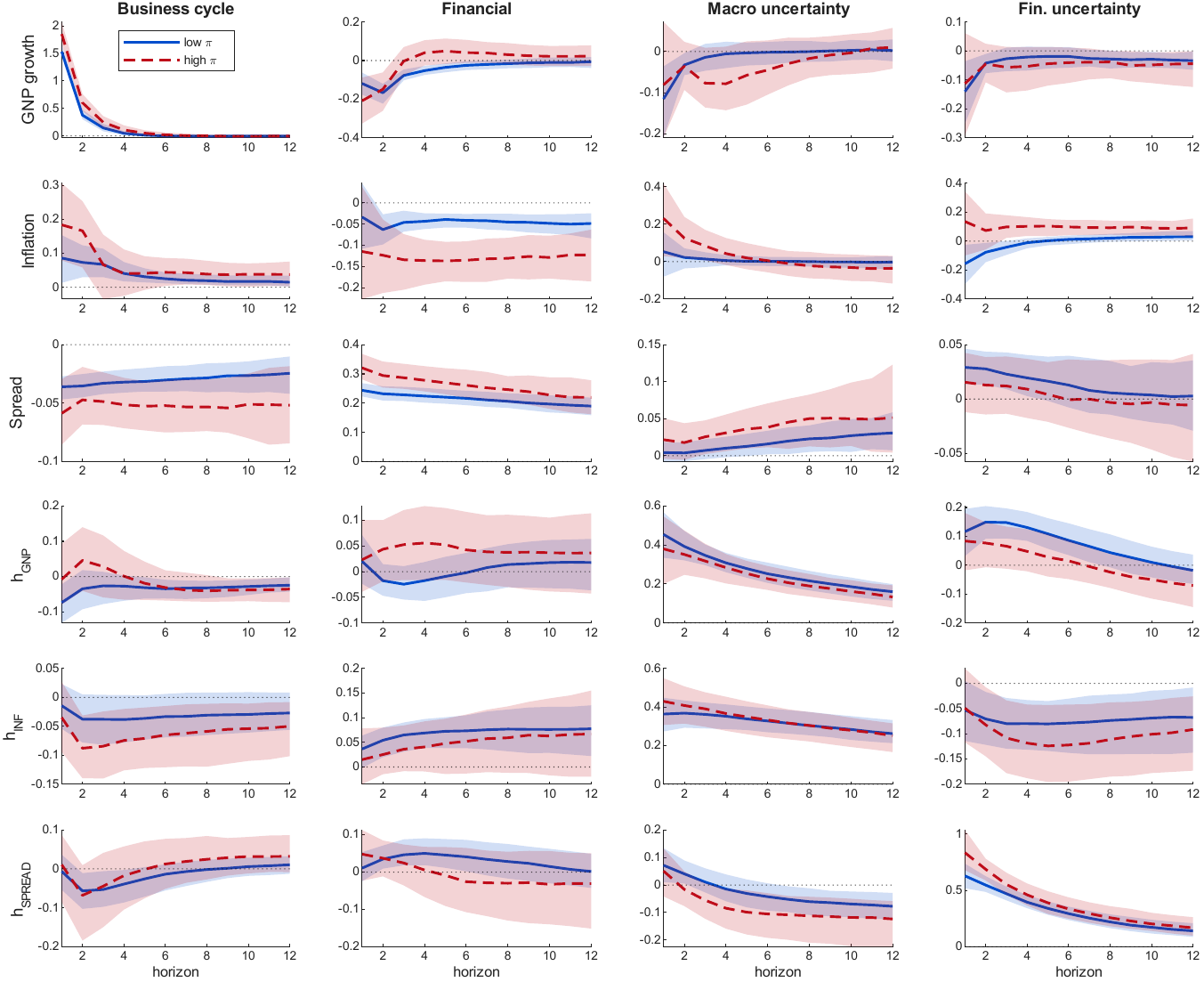}
\caption{\label{fig:girf_regime}State-dependent propagation of identified structural shocks: conditional-mean responses.}
\caption*{\footnotesize \textbf{Notes:} Generalised impulse responses of the conditional mean to one-standard-deviation innovations in the business-cycle, financial, macroeconomic-uncertainty, and financial-uncertainty shocks. Each response is the difference between the mean paths of the shocked and baseline simulations, so it describes the centre of the predictive distribution rather than its tails; the tail responses are in Figure~\ref{fig:risktails}. The top three rows show responses of GNP growth, inflation, and the credit spread; the bottom three rows show the corresponding log-volatilities. Responses use common future innovations in the shocked and baseline paths and are averaged across histories in the low-inflation regime (blue solid lines) and the high-inflation regime (red dashed lines). Shaded areas are $68\%$ credible intervals across posterior draws.}
\end{figure}

\begin{figure}[!htbp]
\centering
\includegraphics[width=\textwidth]{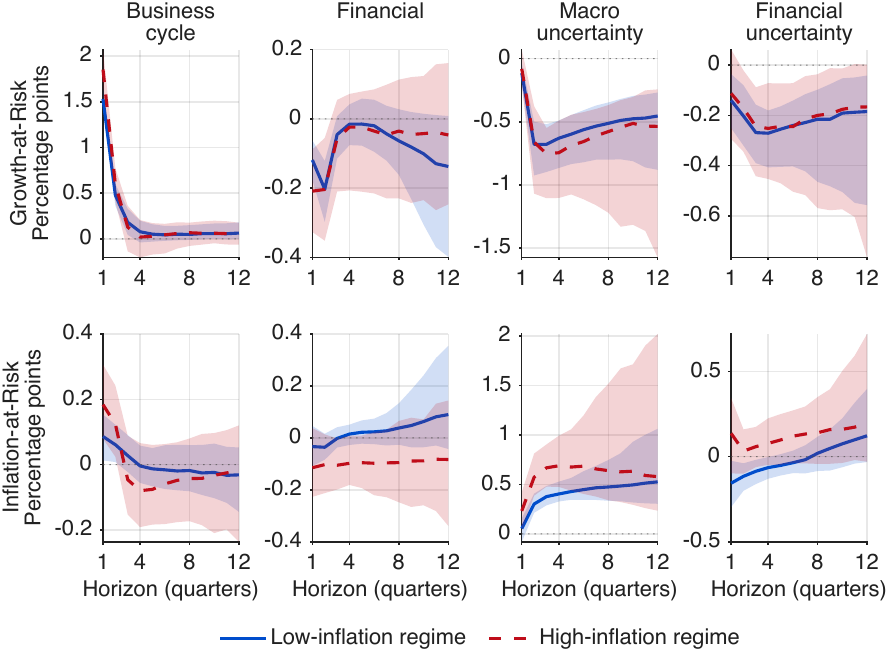}
\caption{\label{fig:risktails}State-dependent propagation of identified structural shocks: tail responses.}
\caption*{\footnotesize \textbf{Notes:} Response of a conditional quantile of the predictive distribution to a one-standard-deviation innovation in each identified shock, by horizon in quarters. The top row reports growth-at-risk, the response of the $5^{th}$ percentile of GNP growth; the bottom row reports inflation-at-risk, the response of the $95^{th}$ percentile of inflation. Quantiles are taken across simulation paths within each posterior draw, so a widening of the predictive distribution is reflected in the response; Figure~\ref{fig:girf_regime} reports the conditional-mean responses to the same impulses. Blue solid lines average responses across histories in the low-inflation regime and red dashed lines across histories in the high-inflation regime; shaded areas are $68\%$ credible intervals across posterior draws.}
\end{figure}

The main text focuses on nonlinear tail-risk responses and their decomposition. This section provides the supporting conventional and one-standard-deviation responses used to establish the economic interpretation of the identified shocks. Figure~\ref{fig:girf_regime} reports responses of the conditional mean: generalised impulse responses of GNP growth, inflation, the credit spread, and their log-volatilities to one-standard-deviation innovations in each of the four shocks, averaged across histories in the low- and high-inflation regimes. Figure~\ref{fig:risktails} then reports the corresponding responses of the tails of the predictive distribution at the same impulse size.

The business-cycle shock moves real activity most strongly, while the financial shock has its largest direct effect on the credit spread. In the posterior medians of the conditional-mean responses, macroeconomic uncertainty has its clearest effects on GNP-growth and inflation volatility, whereas financial uncertainty has its clearest effect on credit-spread volatility. These patterns support the economic labels generated by the max-share criterion. They also show selected state differences in propagation, which are consistent with, but do not identify, the regime dependence in the model's reduced-form level--volatility interactions.

Figure~\ref{fig:risktails} reports the same impulses at the same size, but for the tails of the predictive distribution rather than its centre. At this conventional scale, the uncertainty shocks have the most persistent effect on growth-at-risk. State dependence is clearest for inflation-at-risk: both uncertainty shocks move it more when inflation is high, and financial uncertainty lowers it initially in the low-inflation regime while raising it on impact in the high-inflation regime. These responses motivate the large-shock analysis in the main text.

\clearpage

\end{document}

%% file: tables/tab_modelselection.tex
\begin{table}[!htbp]
\centering
\footnotesize
\caption{\label{tab:modelselection}One-step-ahead joint log predictive scores}
\setlength{\tabcolsep}{8pt}
\renewcommand{\arraystretch}{1.10}
\begin{tabular}{@{}lc@{}}
\toprule
\rowcolor{gray!10}\textbf{Threshold variable} & \textbf{Log predictive score} \\
\midrule
\rowcolor{gray!10}\multicolumn{2}{@{}l}{\textbf{No threshold regimes}} \\
No-threshold model & $-273.1$ \\
\midrule
\rowcolor{gray!10}\multicolumn{2}{@{}l}{\textbf{Two regimes}} \\
GNP growth & $-303.2$ \\
Inflation & $-281.9$ \\
Credit spread & $-293.3$ \\
\midrule
\rowcolor{gray!10}\multicolumn{2}{@{}l}{\textbf{Three regimes}} \\
GNP growth & $-282.9$ \\
Inflation & \textcolor{blue}{\textbf{$-268.4$}} \\
Credit spread & $-306.6$ \\
\bottomrule
\end{tabular}
\caption*{\footnotesize \textit{Notes:} Sum of joint one-step-ahead log predictive scores for GNP growth and inflation over 189 recursive origins, 1975Q1--2022Q1. Higher values indicate a better one-quarter-ahead joint predictive fit for the realised pair. The realisations scored are 1975Q2--2022Q2. Regimes are defined by annual real GNP growth in the GNP-growth specifications, by annual inflation in the inflation specifications, and by the level of the spread in the spread specifications. Bold blue marks the highest score.}
\end{table}

%% file: floats/fig_tailcompare.tex
\begin{figure}[!htbp]
\centering
\includegraphics[width=\textwidth]{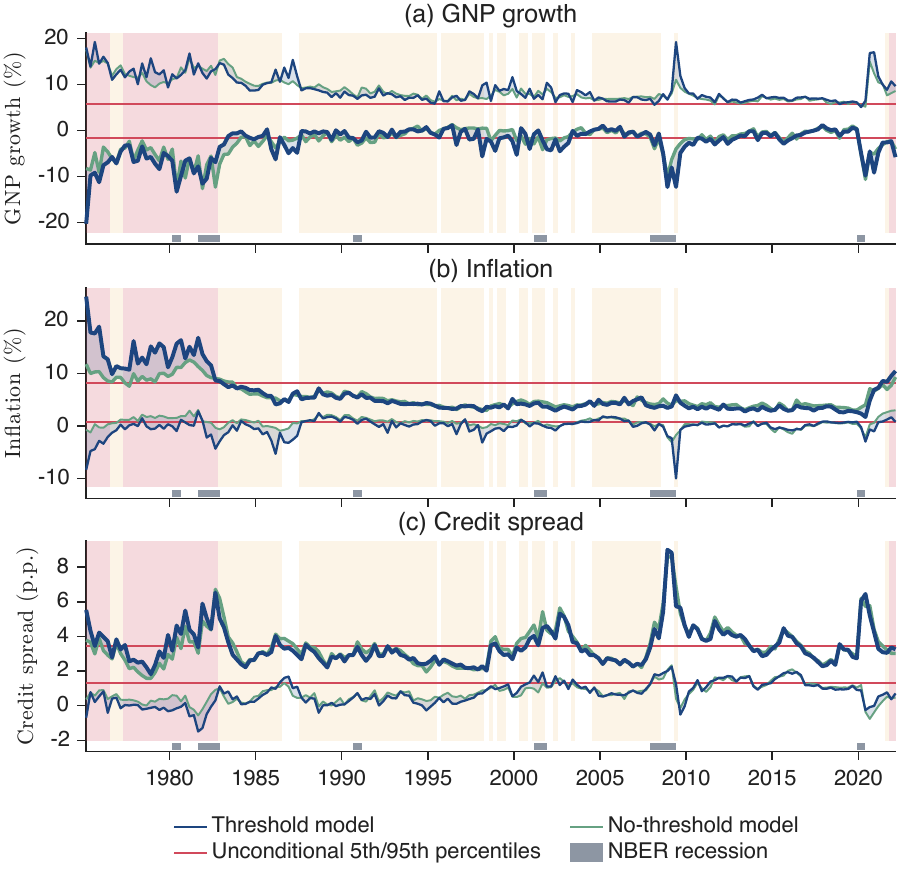}
\caption{\label{fig:tailcompare}Tail risk with and without inflation regimes.}
\caption*{\footnotesize \textbf{Notes:} Four-quarter-ahead 5th and 95th predictive percentiles from recursively estimated threshold (navy) and no-threshold (green) models. GNP growth and inflation are cumulated over the forecast horizon, that is summed over the $h$ quarters following the origin, and the credit spread is measured at its target date. The heavier line denotes the adverse tail. The horizontal red lines are the unconditional 5th and 95th percentiles of the realised series over the same targets, so the distance between a predictive percentile and its reference line measures risk relative to a typical quarter. Background shading reports the modal threshold-model regime at the origin, moderate inflation in amber and high inflation in red; quarters in the low-inflation regime are left unshaded. Grey bars in the lane at the foot of each panel mark NBER recessions, from the peak to the trough quarter. Forecast origins: 1975Q1--2022Q1.}
\end{figure}

%% file: tables/fms_table1.tex
\begin{table}[!htbp]
\centering
\footnotesize
\caption{\label{tab:tailscores}Out-of-sample forecast comparison}
\setlength{\tabcolsep}{4pt}
\renewcommand{\arraystretch}{1.10}
\begin{tabular}{@{}l*{8}{c}@{}}
\toprule
\textbf{Variable} & \multicolumn{8}{c}{\textbf{Forecast horizon}} \\
\cmidrule(l){2-9}
  & \textbf{H1} & \textbf{H2} & \textbf{H3} & \textbf{H4} & \textbf{H5} & \textbf{H6} & \textbf{H7} & \textbf{H8} \\
\midrule
\rowcolor{gray!10}\multicolumn{9}{@{}l}{\textbf{A. Posterior-mean accuracy: RMSE ratio}} \\
GNP growth & \textcolor{blue}{\textbf{0.996}} & \textcolor{blue}{\textbf{0.992}} & \textcolor{blue}{\textbf{0.992}} & 1.004 & 1.002 & 1.005 & 1.010 & 1.007 \\
Inflation & 1.053 & 1.024 & \textcolor{blue}{\textbf{0.999}} & \textcolor{blue}{\textbf{0.971}} & \textcolor{blue}{\textbf{0.954}} & \textcolor{blue}{\textbf{0.939}} & \textcolor{blue}{\textbf{0.930}} & \textcolor{blue}{\textbf{0.920}} \\
Credit spread & \textcolor{blue}{\textbf{0.937}} & \textcolor{blue}{\textbf{0.930}} & \textcolor{blue}{\textbf{0.939}} & \textcolor{blue}{\textbf{0.903}} & \textcolor{blue}{\textbf{0.999}} & 1.003 & 1.003 & 1.006 \\
\midrule
\rowcolor{gray!10}\multicolumn{9}{@{}l}{\textbf{B. Tail-outcome density diagnostic: weighted log-score difference}} \\
GNP growth & \textcolor{blue}{\textbf{27.1}} & \textcolor{blue}{\textbf{43.8}} & \textcolor{blue}{\textbf{102.8}} & $-$11.5 & \textcolor{blue}{\textbf{0.5}} & \textcolor{blue}{\textbf{0.0}} & \textcolor{blue}{\textbf{0.9}} & \textcolor{blue}{\textbf{0.5}} \\
Inflation & $-$0.1 & \textcolor{blue}{\textbf{0.0}} & \textcolor{blue}{\textbf{0.1}} & \textcolor{blue}{\textbf{0.1}} & \textcolor{blue}{\textbf{0.0}} & \textcolor{blue}{\textbf{0.1}} & \textcolor{blue}{\textbf{0.3}} & \textcolor{blue}{\textbf{0.2}} \\
Credit spread & \textcolor{blue}{\textbf{124.9}} & $-$6.6 & $-$0.8 & $-$3.3 & $-$47.3 & $-$16.2 & $-$0.3 & $-$29.1 \\
\bottomrule
\end{tabular}
\caption*{\footnotesize \textit{Notes:} Based on 189 recursive forecast origins. GNP growth and inflation are evaluated as average growth over the horizon, the cumulated change divided by $h$; the spread is the target-date level. Panel A reports threshold/no-threshold RMSE ratios; values below one favour the threshold model. Panel B reports 100 times threshold-minus-no-threshold outcome-weighted log-score differences; positive values favour the threshold model. Bold blue marks values that favour the threshold model. Panel B differences are averages across origins and are dominated by a few extreme quarters, chiefly 2020Q1--2020Q2, when both models assign very little density to the realised outcomes.}
\end{table}

%% file: tables/tab_pit.tex
\begin{table}[!htbp]
\centering
\footnotesize
\caption{\label{tab:pit}Calibration of the one-step-ahead predictive densities}
\setlength{\tabcolsep}{8pt}
\renewcommand{\arraystretch}{1.10}
\begin{tabular}{@{}lcccc@{}}
\toprule
 & \multicolumn{2}{c}{\textbf{Kolmogorov--Smirnov}} & \multicolumn{2}{c}{\textbf{Cram\'er--von Mises}} \\
\cmidrule(lr){2-3}\cmidrule(lr){4-5}
\rowcolor{gray!10}\textbf{Variable} & \textbf{Threshold} & \textbf{No threshold} & \textbf{Threshold} & \textbf{No threshold} \\
\midrule
GNP growth & \textcolor{blue}{\textbf{1.70}}$^{***}$ & 1.90$^{***}$ & \textcolor{blue}{\textbf{0.66}}$^{**}$ & 0.85$^{***}$ \\
Inflation & \textcolor{blue}{\textbf{1.21}}$^{*}$ & 1.44$^{**}$ & \textcolor{blue}{\textbf{0.46}}$^{*}$ & 0.66$^{**}$ \\
Credit spread & \textcolor{blue}{\textbf{1.03}} & 1.18 & \textcolor{blue}{\textbf{0.27}} & 0.31 \\
\bottomrule
\end{tabular}
\caption*{\footnotesize \textit{Notes:} Rossi--Sekhposyan KS and CvM statistics for PIT uniformity from 189 recursive one-step-ahead origins, 1975Q1--2022Q1. Lower values are better; bold blue marks the lower statistic. $^{***}$, $^{**}$ and $^{*}$ denote rejection at the 1, 5 and 10 percent levels, respectively.}
\end{table}

%% file: floats/fig_pit.tex
\begin{figure}[!htbp]
\centering
\includegraphics[width=0.9\textwidth]{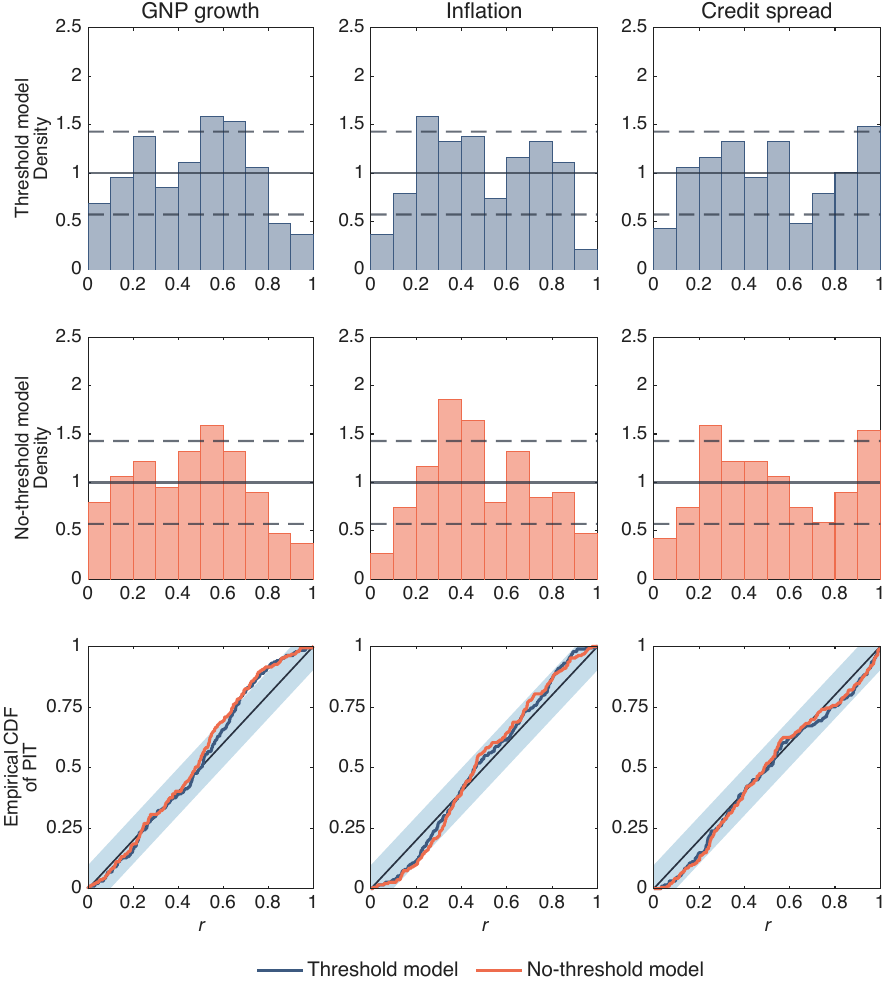}
\caption{\label{fig:pit}PIT calibration of the one-step-ahead predictive densities.}
\caption*{\footnotesize \textit{Notes:} The top and middle rows show PIT histograms for the threshold and no-threshold models, respectively; the solid and dashed lines give the uniform density and its pointwise 95 percent interval. The bottom row plots the empirical PIT CDFs against the 45-degree line; the shaded region is the 5 percent KS band. Recursive one-step-ahead forecasts from origins 1975Q1--2022Q1.}
\end{figure}